\documentclass[traditabstract]{aa}
\usepackage{natbib}
\bibpunct{(}{)}{;}{a}{}{,} 
\usepackage{amsmath}
\usepackage{txfonts}
\usepackage{graphicx}

\begin{document}

\def\tLARR{LARR}
\def\tPPAK{PPAK}
\def\sLARR{\ensuremath{_{\text{\tLARR}}}}
\def\sPPAK{\ensuremath{_{\text{\tPPAK}}}}
\def\cex{\ensuremath{c_{\text{H}\beta}}}
\def\ne{\ensuremath{N_{\text{e}}}}
\def\nee#1{\ensuremath{N_{\text{e},\text{#1}}}}
\def\Te{\ensuremath{T_{\text{e}}}}
\def\teff{\ensuremath{T_{\text{eff}}}}
\def\teffe#1{\ensuremath{T_{\text{eff},\text{#1}}}}
\def\Tee#1{\ensuremath{T_{\text{e},\text{#1}}}}
\def\cbm{\ensuremath{\text{cm}^{-3}}}
\def\sb{\ensuremath{\mbox{erg}\,\mbox{cm}^{-2}\mbox{s}^{-1}\mbox{arcsec}^{-2}}}
\def\sba{\ensuremath{\mbox{erg}\,\mbox{cm}^{-2}\mbox{s}^{-1}\mbox{arcsec}^{-2}\mbox{\AA}^{-1}}}
\def\kms{\ensuremath{\text{km}\,\text{s}^{-1}}}
\def\rCSSP{CSSP03}
\def\rGVM{GVM98}
\def\rMiClWa{MCWH91}
\def\rMiClWao{MCW89}
\def\rPlSo{PS90}
\def\rKwHe{KH98}
\def\rKa{K86}
\def\rSK{SK89}
\def\rMP{MP89}
\def\rJQA{JQA87}

\def\roNGCblue{\object{NGC\,7662}}
\def\roNGCeye{\object{NGC\,6826}}
\def\roNGCowl{\object{NGC\,3587}}
\def\roMtt{\object{M\,2-2}}
\def\roIC{\object{IC\,3568}}
\def\rNGCblue{NGC\,7662}
\def\rNGCeye{NGC\,6826}
\def\rNGCowl{NGC\,3587}
\def\rMtt{M\,2-2}
\def\rIC{IC\,3568}

 \title{Spatially resolved spectroscopy of planetary nebulae\\ and their halos
  \thanks{Based on observations collected at the Centro Astron\'omico Hispano Alem\'an (CAHA), operated jointly by the Max-Planck Institut f\"ur Astronomie and the Instituto de Astrof\'isica de Andalucia (CSIC).}
  \thanks{Figures 24--29 are only available in electronic form via http://www.edpsciences.org.}}
\subtitle{I.\@ Five galactic disk objects}

\author{Christer Sandin \and Detlef Sch\"onberner \and Martin M.\@ Roth \and Matthias Steffen \and Petra B\"ohm \and Ana Monreal-Ibero}
\offprints{Christer Sandin, \email{CSandin@aip.de}}
\institute{Astrophysikalisches Institut Potsdam, An der Sternwarte 16, D-144\,82 Potsdam, Germany}
\date{Received 25 February 2008 / Accepted 20 May 2008}

\abstract{
Strong mass loss off stars at the tip of the asymptotic giant branch (AGB) profoundly affects properties of these stars and their surroundings, including the subsequent planetary nebula (PN) stage. %
With this study we wanted to determine physical properties of mass loss by studying weakly emitting halos, focusing on objects in the galactic disk. Halos surround the, up to several thousand times, brighter central regions of PNe. Young halos, specifically, still contain information of the preceeding final mass loss stage on the AGB. %
In the observations we used the method of integral field spectroscopy with the PMAS instrument. This is the first committed study of halos of PNe that uses this technique. We improved our data analysis by a number of steps. In a study of the influence of scattered light we found that a moderate fraction of intensities in the inner halo originate in adjacent regions. As we combine line intensities of distant wavelengths, and because radial intensity gradients are steep, we corrected for effects of differential atmospheric refraction. In order to increase the signal-to-noise of weak emission lines we introduced a dedicated method to bin spectra of individual spatial elements. We also developed a general technique to part the temperature-sensitive oxygen line [\ion{O}{iii}]\,$\lambda4363$ from the adjacent telluric mercury line Hg$\,\lambda4358$ -- without using separate sky exposures. By these steps we avoided introducing errors of several thousand Kelvin to our temperature measurements in the halo. %
For {\rIC} we detected a halo. For {\rMtt} we found a halo radius that is 2.5 times larger than reported earlier. We derived radially densely sampled temperature and density structures for four nebulae, which all extend from the central regions and out into the halo. {\rNGCblue}, {\rIC}, and {\rNGCeye} show steep radially increasing temperatures and a hot halo, indicating that the gas in the halo is not in thermal equilibrium. {\rMtt} shows a larger temperature in the central region and an otherwise constant value. From the density structures we made estimates of core and halo masses and -- for the first time reliable -- mass loss rates at the tip of the AGB. All four objects show inwards radially increasing mass loss rate structures, which represent a rise by a factor of about 4--7, during the final mass loss phase, that covers a time period of approximately $10^4$ years. Within a factor of two, the average of the maximum mass loss rates, which are distance dependent, is $\dot{M}_{\text{max}}\!\simeq\!10^{-4}\,\text{M}_{\sun}\text{yr}^{-1}$.}%


\maketitle

\section{Introduction}\label{sec:introduction}
Stars on the thermally-pulsing asymptotic giant branch (AGB) are in the final stages of evolution before they turn into cooling white dwarfs. During this phase these stars lose nearly all of the mass outside their core in an increasingly strong stellar wind. The matter of such winds is enriched with nuclearly synthesized elements, which have been dredged up from the interior during the evolution; thereby contributing to the galactic chemical evolution. Models of strong (dust-driven) stellar winds are described by, for example \citet{Bo:88}, \citet{FlGaSe:92}, \citet{WiLBe.:00}, \citet{HoGaArJo:03}, and \citet{CSa:03,Sa:08}. Two recent overviews of mass loss from cool stars are given by \citet{Wi:00} and \citet{Ho:05}. Although there is progress in understanding both stellar evolution and mass loss theoretically, observational details of, in particular, the last phase at the tip of the AGB have remained obscure. The lifetime of the final mass loss stage is short, and selection therefore works against its detection, making detailed observational studies difficult. 

Planetary nebulae (PNe) are in transit between the tip of the AGB and the white dwarf domain of the Hertzsprung-Russell diagram. These extended objects occur in rather large numbers and can readily be observed in the visible wavelength range, where a significant fraction of the luminosity is emitted in a few emission lines. The morphological variation of PNe is large \citep[see, e.g.,][and references therein]{BaFr:02,SaMoSa.:07}. Of those nebulae with a spherical or elliptical shape the central parts consist of a relatively bright rim with a fainter attached shell; both are built-up by an interplay between photo-ionization and wind interaction \citep[see, for example,][]{FrBaRi:90,MaSc:91,Fr:94,Me:94,PeScSt.:04}. Often a faint, and in many cases relatively large, halo surrounds the central PN region; see the PN halo atlas of \citet[hereafter {\rCSSP}]{CoScSt.:03} for examples. Such halos are with the help of models identified as the ionized AGB wind \citep{StSc:03,ViGaMa:02,ViMaGa:02}. Halos of younger nebulae are less affected by PN shaping processes than the central parts are, and therefore contain information about the preceeding mass loss episode; this is not the case with so-called recombination halos of older nebulae \citep[see, e.g.,][]{Ty:86,CoScSt.:00}. Due to very low surface brightnesses little is so far known about physical properties of halos. The purpose of this study is to improve our knowledge by studying halos of a number of PNe in detail.

Integral field spectroscopy (IFS) is a suitable tool for performing spectroscopy on regions spanning very weak to very strong surface brightnesses, such as PNe with halos. Through a generally large field of view IFS permits a co-adding (binning) of any number of spectra of individual spatial elements on an integral field unit (IFU), which covers a defined area on the sky. In this study we used the Potsdam Multi-Aperture Spectrophotometer \citep[PMAS,][]{RoKeFe.:05} in order to perform a plasma diagnostic study of selected PNe, including both central parts and halos. Our goal was to make measurements of electron temperatures and densities, in order to provide spatially resolved radial structures of these properties. While a density allows a determination of mass loss rates, a knowledge of the electron temperature is important when calculating abundances.

We discern between objects in terms of the metalicity, and use four groups of objects: galactic disk, halo, Magellanic Cloud (MC), and other extragalactic PNe. The five objects we selected for this study are all galactic disk nebulae with circular outer regions. {\rNGCblue} and {\rNGCeye} are both well studied objects of considerable size with a well-confined halo. {\rIC} is a closely circular nebula at intermediate distance, that, so far, is reported not to have a halo. {\rMtt} is a small and distant round nebula, that is not often studied. Although a present halo is reported in the literature, the center-to-halo flux ratios are too low \citep[2--5,][hereafter {\rGVM}]{GuViMa:98} to verify the existence of a weak halo (these ratios must be $\ge\!100$, \rCSSP). {\rNGCowl}, finally, is a large evolved nebula with a potential recombination halo. The entire radial extent of the halo was, furthermore, not covered for all objects, but the shell-halo transition region and inner halo were covered in every case. Priority was given to cover the blue wavelength range, including Balmer lines and the temperature sensitive oxygen lines. Some observations were additionally performed in the red wavelength range, in order to cover H$\alpha$ and the density sensitive sulfur lines, where present. It should be noted that only a few emission lines are visible in the halo, where densities normally are too low to be determined using line ratios. 

The IFS method allows a determination of physical properties in two dimensions, but studies presenting two-dimensional maps of PNe are not new. Temperature maps of various nebulae, using different methods of observations are presented by \citet{ReWo:82}, \citet[hereafter {\rJQA}]{JaQuAf:87}, \citet{MaPaPa:01}, \citet{HyMeLe.:01}, \citet{RuBhDu.:02}, and \citet{WeLi:04}. \citet{LiLiLu.:04} and \citet{MoScGr.:04,MoScGr.:05} derive densities and temperatures for two nebulae using a long slit scanning method. \citet[also see \citealt{TsWaPe.:07}]{TsWaPe.:08} measure temperatures, densities and abundances for three nebulae using the FLAMES instrument and the Argus IFU. All these studies focus on the central, brighter, parts of nebulae. In the only study that is similar to ours, so far, \citet{MaViRi.:07}, using PMAS with the PPAK IFU, calculate a temperature map of a knot in the halo of NGC\,6543.

In the following we give the details of the observations in Sect.~\ref{sec:observations}, followed by a description of the performed data reduction in Sect.~\ref{sec:datareduction}, where we also describe how we correct for differential atmospheric refraction (DAR). Before we present the results in Sect.~\ref{sec:results}, we treat important issues of the data analysis in Sect.~\ref{sec:dataanalysis} -- including a discussion on scattered light and descriptions of new dedicated methods to deal with sky subtraction and binning of spatial elements. Our interpretation of the physical results are then found in Sect.~\ref{sec:discussion}. The paper is closed with conclusions in Sect.~\ref{sec:conclusions}.

\section{Observations}\label{sec:observations}
On five different occasions the five objects studied here were observed with the PMAS instrument, that is attached to the 3.5m telescope at Calar Alto. We used the lens array (\tLARR) and {\tPPAK} IFUs \citep{RoKeFe.:05,KeVeRo.:06}. Instrumental configurations, weather conditions, and details on the exposures taken are given in the observational journal in Table~\ref{tab:observations}. In addition to the science exposures continuum and arc lamp flat-fields were taken, as well as spectrophotometric standard-star exposures. In order to correct for varying fiber-to-fiber transmission sky flat-fields were taken for all targets, but {\rNGCeye}, where a normalized continuum lamp flat-field was used instead. Note that in most cases no separate observations of the sky -- completely void of halo emission lines -- were made; instead an alternative method was adopted where a set of telluric lines are fitted together with the object lines, cf.\ Sect.~\ref{sec:dataanalysisskysub}.  Data of {\rNGCowl} from 2004 were sampled using the beam switch technique (BSW, cf.\@ Sect.~\ref{sec:dataanalysisskysub}). This data, however, did not permit an accurate detection of the weak auroral line [\ion{O}{iii}]$\,\lambda4363$, and the object was therefore observed again in 2006, then without using the BSW technique.

\begin{table*}
\caption{Journal of observations}\label{tab:observations}
\begin{tabular}{lcrcclcl}
\hline\hline
\noalign{\smallskip}
Object           & Date       & \multicolumn{1}{l}{Grating/}      & Wavelength & IFU$^{\mathrm{a}}$                & \multicolumn{2}{c}{Weather conditions} & Exposures$^{\mathrm{c}}$\\
\cline{6-7}
                 & Observed   &  \multicolumn{1}{r}{Dispersion/R} & Range      &                    & \multicolumn{1}{c}{Sky Quality$\!$} & $\!$Seeing$^{\mathrm{b}}$    &    \\
                 &            & -  [\AA/pixel]  -                 & [\AA]      &                    &             &                          & [seconds (s) or minutes (m)]\\
\noalign{\smallskip}
\hline
\noalign{\smallskip}
{\roNGCowl}  & 2004.02.12 & V300/ 1.67/$\phantom{0}800$       & 3590--7000  & 1\farcs0\,$^{\mathrm{BSW}}\sLARR$ & photometric & 1\farcs1--1\farcs4                  & $20^{\mathrm{m}\,:\,1.20}_{94\mathrm{W}\,::\,\mathrm{a}}$, $20^{\mathrm{m}\,:\,1.05}_{110\mathrm{W}\,::\,\mathrm{b}}$, $20^{\mathrm{m}\,:\,1.06,1.10}_{126\mathrm{W}\,::\,\mathrm{c}}$\\[0.7ex]
            & 2004.02.15 & V1200/0.38/2690                   & 3640--4400 & 1\farcs0\,$^{\mathrm{BSW}}\sLARR$ & photometric & 1\farcs5-2\farcs0                  & $20^{\mathrm{m}\,:\,1.18}_{94\mathrm{W}\,::\,\mathrm{a}}$, $20^{\mathrm{m}\,:\,1.10}_{110\mathrm{W}\,::\,\mathrm{b}}$, $20^{\mathrm{m}\,:\,1.05}_{126\mathrm{W}\,::\,\mathrm{c}}$\\[0.7ex]
            & 2006.04.26 & V1200/0.35/5080                   & 6210--6870  & 1\farcs0\,\sLARR & not optimal & 1\farcs1--1\farcs2                  & $20^{\mathrm{m}\,:\,1.05,1.05}_{74\mathrm{W}\,::\,1}$, $20^{\mathrm{m}\,:\,1.06,1.08}_{90\mathrm{W}\,::\,2}$\\[0.7ex]
            &            &                                   &            &                  &             & 0\farcs9--1\farcs1                 & $20^{\mathrm{m}\,:\,1.10,1.12,1.17}_{106\mathrm{W}\,::\,3}$, $20^{\mathrm{m}\,:\,1.21,1.29}_{122\mathrm{W}\,::\,4}$\\[0.7ex]
            & 2006.04.28 & V1200/0.35/3230                   & 4300--5040  & 1\farcs0\,\sLARR & very good   & 1\farcs5--2\farcs0                  & $20^{\mathrm{m}\,:\,1.07,1.09,1.13}_{90\mathrm{W}\,::\,2}$, $20^{\mathrm{m}\,:\,1.16,1.23}_{106\mathrm{W}\,::\,3}$, $30^{\mathrm{m}\,:\,1.37}_{122\mathrm{W}\,::\,4}$\\[0.7ex]
{\roIC}      & 2004.02.16 & V300/ 1.67/$\phantom{0}800$       & 3590--7000  & 1\farcs0\,$^{\mathrm{BSW}}\sLARR$ & photometric & $\overline{1\farcs1}$                    & $5^{\mathrm{m}\,:\,1.42}_{\mathrm{c}\,::\,0}$, $20^{\mathrm{m}\,:\,1.42}_{\mathrm{c}\,::\,1}$\\[0.7ex]
{\roNGCblue} & 2005.09.09 & V600/ 0.81/1340                   & 3490--5150  & 1\farcs0\,\sLARR & not optimal & 1\farcs0--1\farcs7                  & $15^{\mathrm{s}\,:\,1.32}_{\mathrm{c}\,::\,1}$, $200^{\mathrm{s}\,:\,1.28}_{16\mathrm{W}\,::\,2}$, $30^{\mathrm{m}\,:\,1.11,1.17}_{32\mathrm{W}\,::\,3}$\\[0.7ex]
{\roNGCeye}  & 2006.10.28 & V1200/0.35/3210                   & 4270--5010  & $2\farcs68$\,\sPPAK& very good   & 1\farcs6                      & $30^{\mathrm{s}\,:\,1.06}_{\mathrm{c}\,::\,1}$, $5^{\mathrm{m}\,:\,1.08}_{\mathrm{c}\,::\,1}$, $30^{\mathrm{m}\,:\,1.09}_{30\mathrm{SE}\,::\,2}$\\[0.7ex]
{\roMtt}     & 2007.10.04 & V600/ 0.81/1360                   & 3550--5200  & 0\farcs5\,\sLARR & photometric & 1\farcs3--1\farcs5                  & $10^{\mathrm{m}\,:\,1.16,1.16,1.16}_{\mathrm{c}\,::\,1}$, $45^{\mathrm{m}\,:\,1.11,1.08}_{8\mathrm{W}\,::\,2}$, $45^{\mathrm{m}\,:\,1.07}_{16\mathrm{W}\,::\,3}$\\[0.7ex]
            & 2007.10.05 & R600/ 0.78/1980                   & 5370--6960  & 0\farcs5\,\sLARR & photometric & 1\farcs0--1\farcs3                      & $10^{\mathrm{m}\,:\,1.07,1.06,1.06}_{\mathrm{c}\,::\,1}$, $45^{\mathrm{m}\,:\,1.06}_{8\mathrm{W}\,::\,2}$, $45^{\mathrm{m}\,:\,1.08}_{16\mathrm{W}\,::\,3}$\\
\noalign{\smallskip}
\hline
\noalign{\smallskip}
\end{tabular}\\
{\scriptsize $^{\mathrm{a}}$ The number indicates the size of each spatial element. The subscript indicates what IFU was used, and a superscript indicates whether beam switching (BSW) was used; cf.\ Sect.~\ref{sec:dataanalysisskysub}.}\\
{\scriptsize $^{\mathrm{b}}$ The seeing is estimated from guide star measurements in the R-band.}\\
{\scriptsize $^{\mathrm{c}}$ The superscript index ($x\,:\,x,x,...$) indicates minutes (m) or seconds (s), followed by the airmass of the respective exposure of a particular pointing, the subscript index ($x\,::\,x,...$) indicates the location (specified in arcseconds) of the center of the IFU relative to the position of the central star (c), followed by the tile number/letter used in Fig.\ref{fig:obsimages}.}
\end{table*}

The {\tLARR} IFU holds $16\!\times\!16$ separate fibers, where each fiber represents a spatial element, or a so-called \emph{spaxel}, on the sky. In the 1\farcs0 (0\farcs5) sampling mode every {\tLARR} pointing (IFU) covers an area of $16\arcsec\!\times\!16\arcsec\!=\!256\arcsec^2$ ($8\arcsec\!\times\!8\arcsec\!=\!64\arcsec^2$) on the sky. In comparison the {\tPPAK} IFU holds 331 fibers, which are arranged in a hexagonal grid \citep[see fig.~5 in][]{KeVeRo.:06}, and cover an area on the sky of about $1\farcm0^2$. Any number of spaxels can be co-added to create a final spectrum (cf.\ Sect.~\ref{sec:dataanalysisspxmap}).

\begin{figure*}
\centering
\includegraphics{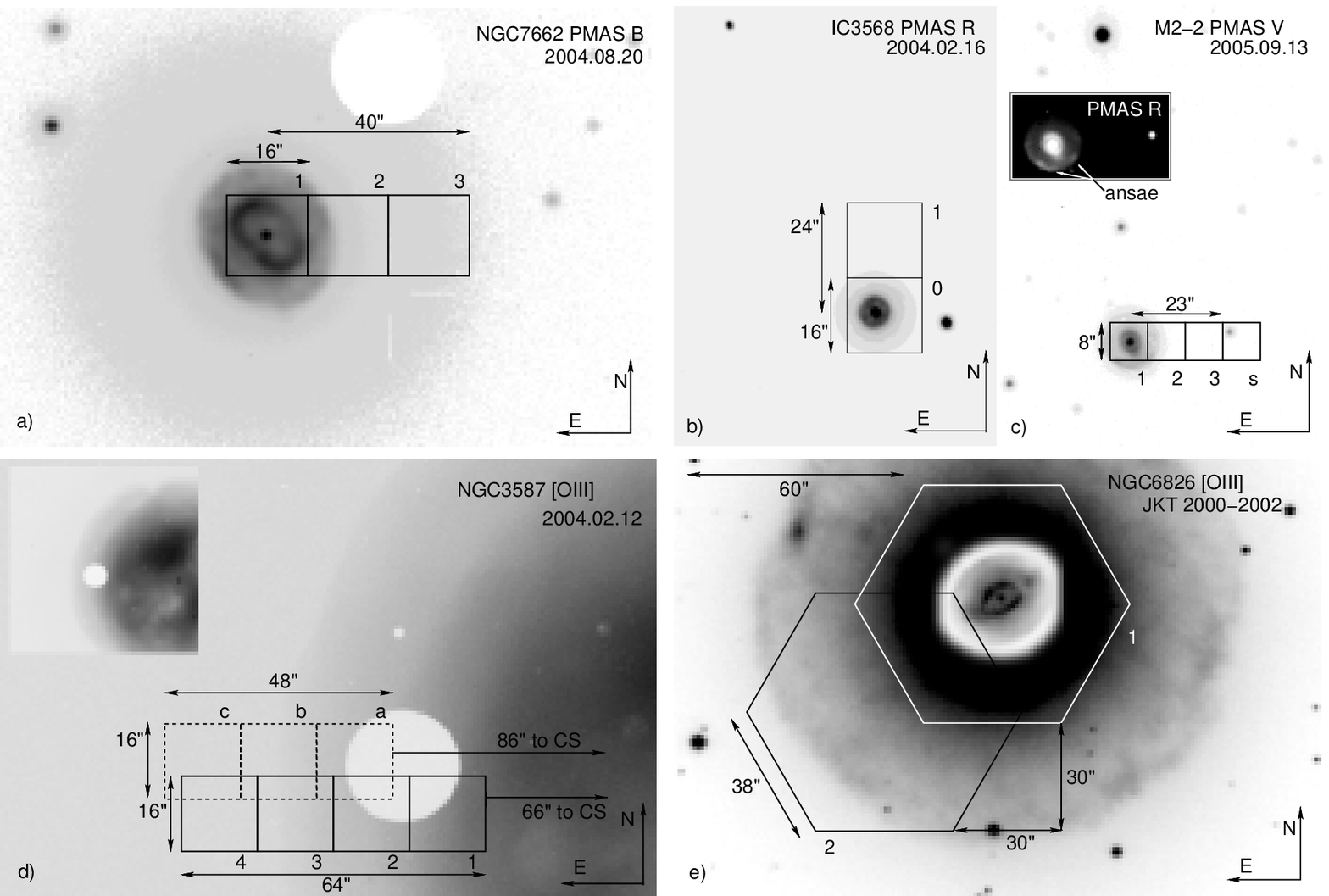}
\caption{Object images illustrating where observations were done. The five panels show: {\bf a)} {\rNGCblue}, {\bf b)} {\rIC}, {\bf c)} {\rMtt}, {\bf d)} {\rNGCowl}, and {\bf e)} {\rNGCeye}. The images in panels {\bf a}--{\bf d} were obtained with the PMAS Acquisition and Guiding (A\&G) camera, and are reproduced in the same scale. The image of {\rNGCeye} in panel {\bf e} was observed with JKT and is taken from the image atlas of {\rCSSP}. The bright disk visible in panels {\bf a} and {\bf d} corresponds to the diaphragm of the A\&G pick-off mirror. The numbered squares/hexagons outline different positions of the PMAS IFUs across the respective PN shell and the halo. In panel {\bf c} the inset shows a corresponding A\&G R-filter image emphasizing the ansae in the southernmost part of the shell; cf.\ Sect.~\ref{sec:resM22}. In panel {\bf d} the tiles drawn with dashed lines were observed on 12.02.2004 and 15.02.2004, and the tiles drawn with solid lines on 26.04.2006 and 28.04.2006; cf.\ Table~\ref{tab:observations}.}
\label{fig:obsimages}
\end{figure*}

We show images of the objects we studied in Fig.~\ref{fig:obsimages}, where the locations of individual IFU pointings are indicated. These images can, with the exception of {\rMtt} and {\rIC}, be compared with the respective image in {\rCSSP} (the image of {\rNGCeye} is taken from this reference).

Wavelength regions and gratings were chosen to cover emission lines crucial when making a plasma diagnostic. All objects were observed in the wavelength region covering the oxygen lines [\ion{O}{iii}]\,$\lambda4363$, $\lambda4959$ and (in most cases) $\lambda5007$, and the Balmer lines H$\gamma$ and H$\beta$. Some of the other emission lines in this blue part of the spectrum, observed simultaneously, were [\ion{Ne}{iii}]\,$\lambda3869$, H$\delta$, \ion{He}{i}\,$\lambda4471$, \ion{He}{ii}$\,\lambda4686$, and the density sensitive doublet [\ion{Ar}{iv}]\,$\lambda\lambda4711,\,4740$. Note, however, that [\ion{Ar}{iv}]\,$\lambda4711$ is blended by \ion{He}{i}\,$\lambda4713$, which can make the use of this line ratio uncertain (see sect.~\textsc{ii} in \citealt{StKa:89}, hereafter {\rSK}, and sect.~2 in \citealt{WaLiZh.:04}). Moreover, [\ion{Ne}{iii}]\,$\lambda3968$ blended with H$\epsilon$ was detected in every case when  [\ion{Ne}{iii}]\,$\lambda3869$ was observed, but was not considered further. A resolved density sensitive oxygen doublet [\ion{O}{ii}]\,$\lambda\lambda3726,\,3729$, requiring a separate setting with a high resolution grating, was only observed for two of the five objects. In the red part of the spectrum the goal was to cover H$\alpha$, and the nitrogen lines [\ion{N}{ii}]\,$\lambda\lambda6548$,\,$6583$, along with the density sensitive sulfur doublet [\ion{S}{ii}]\,$\lambda\lambda6717,\,6731$.

\section{Data reduction}\label{sec:datareduction}
The tools we used in the data reduction, \textsc{P3d\_online} and \textsc{Ppak\_online}, are parts of the PMAS \textsc{P3d} pipeline \citep[\citealt{RoKeFe.:05}]{TBe:02}. We first describe the data reduction method for observations carried out with the {\tLARR} IFU.

At first the bias level was subtracted and cosmic-ray hits removed. Second, a trace mask was generated from an internal continuum calibration lamp exposure, identifying the location of each spectrum on the CCD along the direction of cross-dispersion. Third, a dispersion mask for wavelength calibration was created using an arc-exposure. In order to minimize effects due to a significant flexure in the instrument all continuum and arc lamp exposures were taken within two hours of the respective science exposure. Fourth, a correction to the fiber-to-fiber sensitivity variations was applied by dividing with an extracted and normalized sky or continuum lamp flat-field exposure. In this process the data was changed from a CCD-based format to a row-stacked-spectra (RSS) format. Fifth, the data was corrected for differential atmospheric refraction (DAR), using the procedure outlined below. As a final step flux calibration was performed in \textsc{iraf} using standard-star exposures.

The reduction procedure is slightly different with PPAK data. Apart from a different number of spaxels and a different spaxel geometry, special care was taken to compensate effects of instrument flexure. Additional calibration fibers included in the {\tPPAK} fiber bundle were illuminated with a Thorium/Argon lamp, producing single spots on the CCD. The locations of these spots shift with flexure in the same way as the measured spectra and can therefore be used to calculate shifts between calibration and science exposures. After fixing the so-called shift boxes the steps of the data reduction are the same as for the {\tLARR} IFU, i.e.\@, make a bias, generate a trace mask from a continuum (halogen) lamp exposure, generate a dispersion mask from an arc-lamp exposure (mainly He+Hg), and generate a flat-field either from dome flat or continuum lamp exposures; we used a continuum lamp exposure. Afterwards the spectra were extracted from the science exposures by applying the prepared masks/frames with respect to the calculated shift boxes. The result is stored in the RSS format.

In several cases, and in particular where there was only one science frame, cosmic-ray hits were removed using the \textsc{L.A.\@ Cosmic} routine \citep{vDo:01}. Since this routine is adapted for images and not spectra, the input parameters had to be chosen carefully to ensure that ideally no, or as few spectral lines as possible, were removed together with the cosmic-ray hits. Typical values used with the routine were: $f_{\text{lim}}\!=\!30$--$50$ and $\sigma_{\text{clip}}\!=\!4$--$12$; the most careful (large) numbers were required with the standard star raw data. Any remaining hits at wavelengths close to emission lines were then manually removed using \textsc{splot} in \textsc{iraf} on the final RSS-data -- before correcting for DAR or flux calibrating.

In order to capture all of the standard star flux, we summed as many spaxels as possible. These summed spectra were then sky subtracted using an averaged sky spectrum, before creating the respective sensitivity function. This is not a problem since these spectra typically show few emission lines. Unlike the square lenslets, which are not separated on the {\tLARR} IFU, the fibers on the {\tPPAK} IFU are circular, of 2\farcs68 diameter, and are separated by 3\farcs6. Making an absolute flux calibration in this case, the total flux of the standard star has to be corrected for light lost between the fibers. We made such a correction by first fitting the intensity and position of a two-dimensional Gaussian to the standard star exposure data. Thereafter the fraction of light illuminating the area between fibers was estimated, taking the shape of the Gaussian curve into account.

Several of the objects were observed at such a high airmass that effects of DAR become visible -- i.e., a wavelength-dependent spatial shifting of spectral images on the IFU. Other factors, which also determine the extent of effects of DAR are the geometrical size of the spaxels, and the presence of strong intensity gradients. In effect, the smaller the spaxels the more of the flux of different wavelengths end up on separate spaxels. It is important to correct for DAR when comparing lines of different wavelength. In order to do the correction, and (re)place flux from the same spatial region in the same spaxel, the approach of \citet{Fi:82} is used to calculate an image offset vector for each wavelength bin. Each image, at every wavelength, is thereafter shifted with respect to a pre-defined reference wavelength (that mostly is set to $\lambda_{\text{ref}}\!=\!5050\,${\AA}), using a (fractional) bilinear interpolation procedure. Doing this the intensity in each spaxel becomes a function of four spaxels -- introducing an unavoidable smoothing, which additionally makes the intensity error estimation more challenging. In contrast to our approach \citet{ArMeGa:99} use shifting positions of an intensity maximum at discrete wavelengths in the data to calculate the offset vector. Their approach is difficult to apply when a maximum is located to the edge of the IFU, which for example is the case when the IFU is placed in the halo region where intensities decrease monotonically.

An additional consequence of the DAR correction procedure is that those strips of spaxels on the two edges of a (rectangular) IFU have to be masked, which for some wavelength bins depend on flux from regions outside the IFU. The width of the strips on each side, and on which side they have to be masked, depend on the airmass during the observations, the wavelength, and the location on the sky; see Figs.~\ref{fig:resspxmapNGC7662} and \ref{fig:resspxmapM22} for examples. Note that no correction for DAR was made for {\rNGCeye}, that was observed with {\tPPAK} at a rather low airmass. The large and separated spaxels of this IFU make the usefulness of the interpolation, as described above, doubtful.

\section{Data analysis -- issues critical to results}\label{sec:dataanalysis}
Before presenting final results in Sect.~\ref{sec:results} we next address four topics that all have an important influence on properties of the outcome. We first discuss the influence of scattered light, estimated with a measured point spread function, in Sect.~\ref{sec:dataanalysisstray}. Then we present a method for dealing with telluric (i.e., sky emission) lines in Sect.~\ref{sec:dataanalysisskysub}. This is followed by the presentation of a dedicated method for binning spectra to increase the signal-to-noise in Sect.~\ref{sec:dataanalysisspxmap}, and a discussion of dereddening in Sect.~\ref{sec:dataanalysisdered}.

\subsection{Estimating the influence of scattered light}\label{sec:dataanalysisstray}
In the halo of a planetary nebula the surface brightness in any line can be more than a thousand times weaker than in the rim, where most flux is emitted. An instrumental point spread function (PSF) of a single source can be measured out to very large radial distances, where the relative intensity is $10^5$ times as weak as in the center \citep{Ki:71}. In their work on halos of PNe, \citet{MiClWa:89b} and {\rCSSP} (sect.~4.1) study effects of scattered light. A qualitative study is motivated also for us to see where emission in the halo region originates, because all instruments scatter light to some degree.

\begin{figure*}
\sidecaption
\includegraphics{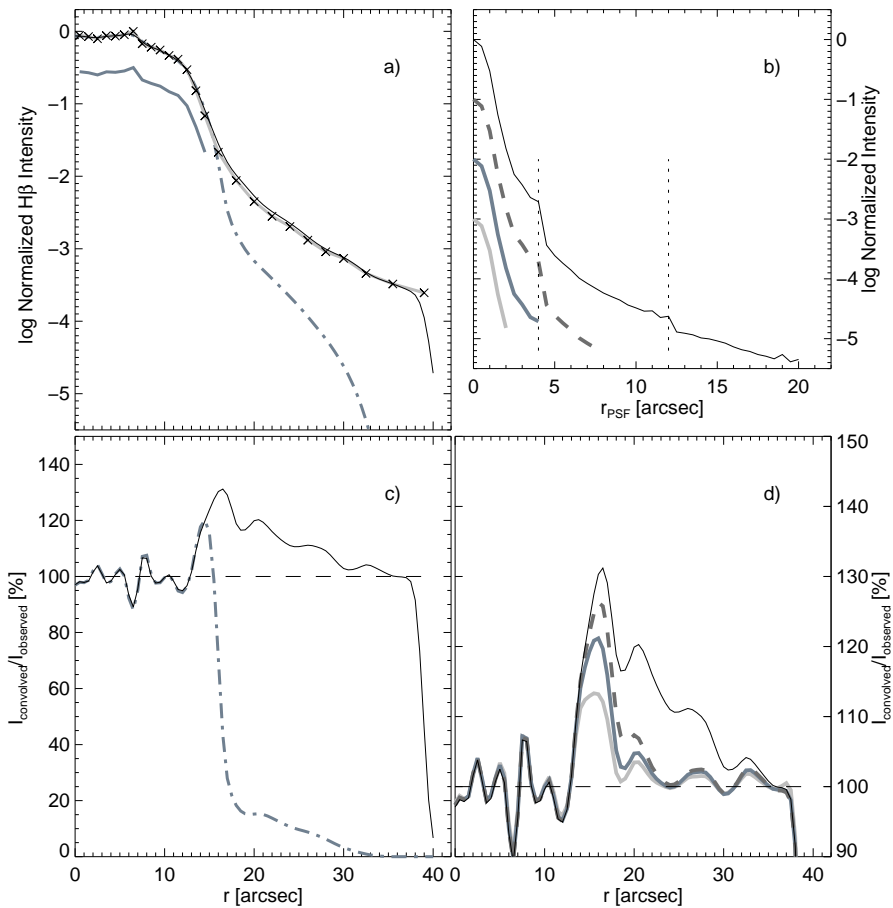}
\caption{Estimating effects of scattered light in the observed data. Panel {\bf b} shows the instrumental point spread function (PSF) measured for the bright star Vega, out to $r\!=\!20{\arcsec}$ away from the PSF center. The ordinate is logarithmic. The three profiles that are vertically offset, by -1\,dex incrementally, show the PSF truncated at: $r_{\text{PSF}}\!=\!8\arcsec$ (dashed line), $r_{\text{PSF}}\!=\!4\arcsec$ (gray line), and $r_{\text{PSF}}\!=\!2\arcsec$ (light gray line). The two tile boun\-daries are indicated with vertical dotted lines. The observed intensity structure of H$\beta$, that is convolved with the PSF, is illustrated in panel {\bf a} (solid black line), again with a logarithmic ordinate. The observed normalized intensity structure is indicated with crosses and the light gray line. For a study of the appearance of a non-existent halo the structure was truncated at $r\!=\!15\arcsec$ (shown by the gray line that is vertically offset by -0.5\,dex), and then convolved with the PSF. The resulting structure is shown with a dash-dotted line. Convolved-to-observed ratios of the same two structures are shown in panel {\bf c}. Panel {\bf d}, finally, shows how the convolved structure plot depends on the radial extent of the PSF. The thin black line uses the full PSF (as in panel {\bf c}). For the remaining three lines we used a truncated PSF -- the colors of these lines are the same as in panel {\bf b}. The figures show that in the inner halo the contribution of scattered light mostly comes from adjacent regions (gray lines; {\bf d}). For further details see Sect.~\ref{sec:dataanalysisstray}.}
\label{fig:dataanalysisstray}
\end{figure*}

In order to determine the instrumental PSF we first observed the bright star Vega ($\alpha$Lyrae) in the red part of the spectrum, across three spatially offset tiles (with consecutively increasing exposure times), using the {\tLARR} IFU in the $0\farcs5$ setting. The maximum distance from the central star, where we measured an intensity, was thereby $r\!=\!20\arcsec$. We extracted a radial slice, one spaxel wide (0\farcs5), covering all three tiles at the wavelength of H$\alpha$, and then used it to create an axi-symmetric two-dimensional PSF kernel of 40{\arcsec} diameter. The one-dimensional PSF is illustrated in Fig.~\ref{fig:dataanalysisstray}b; the measured full width at half maximum of the PSF is about $1\farcs4$--$1\farcs8$. Note the jump in intensity at the transition between the two inner tiles -- the amount of scattered light decreases by about 0.6\,dex when Vega is outside the field of view, i.e.\@ outside the lens array and the diaphragm of the A\&G camera pick-off mirror, cf.\@ Fig.~\ref{fig:obsimages}. In this context it should be mentioned that the most accurate PSF -- which would extend across several tiles -- ought to be determined for each spaxel individually, otherwise tile-to-tile jumps are placed incorrectly. Our aim was, however, to make a qualitative estimate of the influence of the PSF and we did not consider such a detailed treatment.

Next we created a two-dimensional intensity structure, of $80\arcsec$ diameter, using the radial surface brightness of {\rNGCblue} in H$\beta$ (cf.\ Sect.~\ref{sec:resNGC7662sb}; the seeing was about the same as during the night the PSF was observed). This structure was then convolved with the PSF kernel in order to achieve a first order estimate of effects of scattered light. Note that in reality the observed intensity structure must be understood as the convolution of the true surface brightness distribution with the same PSF. We show a radial slice of the intensity before and after convolution in Fig.~\ref{fig:dataanalysisstray}a. The convolved-to-observed intensity ratio is shown in Fig.~\ref{fig:dataanalysisstray}c, indicating the difference more clearly. While differences are small inwards of the shell-halo transition (i.e.\@ for $r\!\la\!15\arcsec$), there is a contribution of up to 30\% to the intensity in the outer region from the central regions of the nebula (also see below).

As a test, to see what the intensity structure of a non-existent halo would appear like we then truncated the two-dimensional observed intensity structure at $r\!=\!15\arcsec$, before convolving with the PSF kernel. The result, that is equivalent to a PN without a halo, is illustrated with the dash-dotted lines in Figs.~\ref{fig:dataanalysisstray}a,c. Compared to the solid line the intensity decreases sharply for $15\arcsec\!<\!r\!<\!18\arcsec$, is $\la\!15$\% of its strength, and decreases for $r\!\ge\!18\arcsec$. This demonstrates that the observed halo is real, because the observed data does not show such a steep decline. Likewise this test also shows that a fraction of the emitted intensity in the halo originates in the central regions; this fraction is the largest in the inner halo.

In order to study to what extent different parts of the PSF contribute to the emitted intensity in the halo we made three tests where the PSF is truncated at increasing radii. Figures~\ref{fig:dataanalysisstray}b and d show the result when such a truncation was made at $r_{\text{PSF}}\!=\!2\farcs0$, $r_{\text{PSF}}\!=\!4\farcs0$, and $r_{\text{PSF}}\!=\!8\farcs0$. Evidently local regions contribute more to the emitted intensity in the halo for $15\arcsec\!\le\!r\!\le\!18\arcsec$, where about two thirds of the scattered light comes from the inner $4\arcsec$ of the PSF; and the rest from the remaining PSF. In the outer region, for $18\arcsec\!\le\!r\!\le\!30\arcsec$, it is in contrast the outer part of the PSF, beyond $r_{\text{PSF}}\!=\!8\arcsec$, that contributes the most; the intensity is there $\la\!10\%$ larger because of scattered light from the outer wing of the PSF.

Radial density structures, which are calculated using the Abel transform on intensities (cf.\@ Sect.~\ref{sec:results}), are affected by scattered light. Estimating the influence we first calculated an approximative deconvolved two-dimensional intensity structure of H$\beta$ -- instead of using a forwards convolution; by matching the convolution of the deconvolved structure and the PSF with the observed intensity structure we could determine it by iteration. The resulting density structure, that is based on this deconvolved structure, was found to be up to about 30 per cent less dense than the density structure, that we derived from the observed intensity structure (in the inner halo). This error estimate for the density is directly comparable to the amount of scattered light that we found in the intensity seen in Fig.~\ref{fig:dataanalysisstray}c.

In comparison to these tests it is more difficult to see what the influence is on properties which depend on line ratios. We therefore also convolved the radial surface brightness structures of oxygen [\ion{O}{iii}]$\,\lambda4363$ and $\lambda5007$, and found that the resulting temperature structure is lower by $\la\!6$\%, compared to the measured temperatures. The largest difference is seen in the halo. This contribution to the error of the temperature is of a similar magnitude as that due to a badly known extinction coefficient in the halo (cf.\@ Sect.~\ref{sec:resNGC7662T}).

Consequently, because of scattered light the observed surface brightness of any line is about $0$--$30$\% brighter than the true brightness, and the deviation is the largest in the inner halo, where gradients are the steepest. For line ratios of a particular region -- and quantities which depend on these ratios, such as the electron temperature -- this implies that actual emission is mixed with emission from regions of different physical properties. True extrema in these regions are thus even more pronounced than what is observed. Since the differences are reasonably small, however, we did not correct our measurements for scattered light, but note that such a component is present. Furthermore, for objects which are small (compared to the extent of the PSF) the light contamination in the halo can be expected to be larger, in particular when the entire PN, with halo, fits on one IFU. In order to find out to what degree such observations are affected it is possible to use a procedure similar to the one carried out above.

\subsection{Sky subtraction}\label{sec:dataanalysisskysub}
An important issue, especially when working with weak lines and a relatively low resolution, is the subtraction of telluric (sky) emission lines from the object data. This is of special concern here since no dedicated frames of the sky were sampled, which have a high enough signal-to-noise in each spaxel (with exception of those exposures that are observed in BSW mode). Specifically it is the telluric Hg\,$\lambda4358$ and the -- in many circumstances very -- weak [\ion{O}{iii}]\,$\lambda4363$ lines that are of main interest; this oxygen line is crucial when determining the electron temperature. We considered two approaches of sky subtraction.

\begin{figure}
\centering
\includegraphics{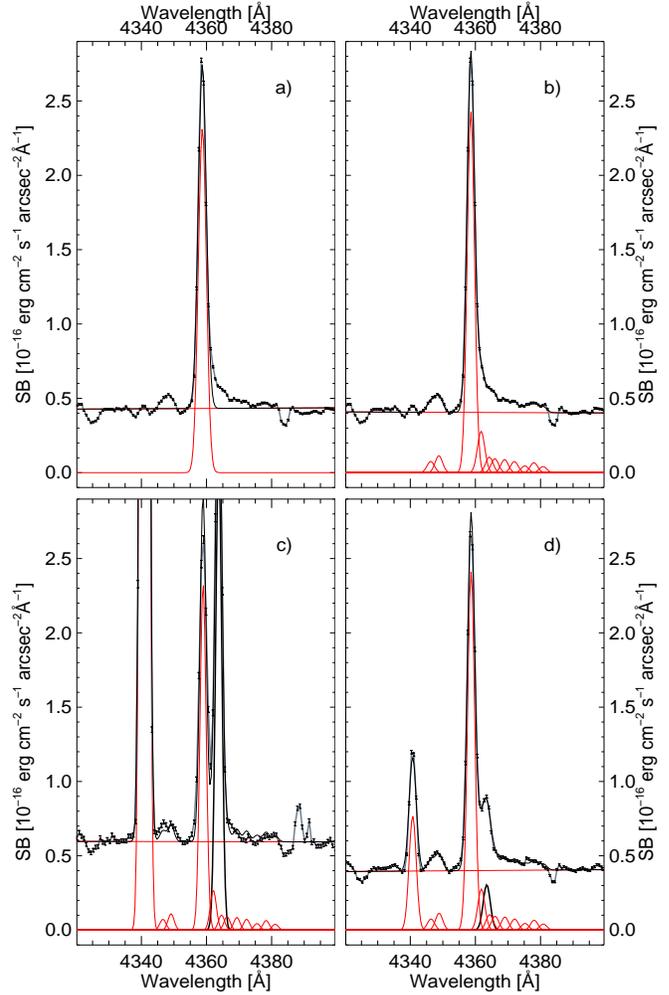}
\caption{Illustration of the sky subtraction procedure described in Sect.~\ref{sec:dataanalysisskysub}, using data of {\rNGCeye}. The four panels show the surface brightness (SB) in the wavelength range 4320--4400\,{\AA} of: {\bf a)} a sky spectrum with a line fit of Hg\,$\lambda$\,4358, {\bf b)} like {\bf a}, but with ten additional lines that together form a set of 11 lines which purpose is to fit the sky emission profile, {\bf c)} strong object lines together with the previously fitted set of 11 sky emission lines, and {\bf d)} that is like {\bf c}, but in the much weaker halo. Lines starting from SB$=\!0.0\,\sba$ show individual fits. In panels {\bf c} and {\bf d} the fit drawn in black is the line [\ion{O}{iii}]$\,\lambda4363$. A total line fit is drawn on top of the observed data. Error bars are shown, but they are very small.}
\label{fig:dataanalysissky}
\end{figure}

In the first approach an averaged sky spectrum, which is sampled in regions where no halo emission lines are found, can be subtracted from all target spectra. On the downside such a sky subtraction is stiff and does not account for minor differences in line shape and wavelength calibration across the IFU. Consequently, for strong telluric lines residuals of ``negative'' emission can remain on one side of the line, and a residual of reverse sign on the other side. Such residuals in many cases prevent a correct estimate of intensities of adjacent weak lines. In comparison, a non-binned sky subtraction requires a separate sky spectrum exposure for each IFU pointing, which is observed using the same exposure time as the object frame in order to not introduce additional noise. This is an alternative but also very time consuming operation. As is the nod-shuffle (BSW) technique \citep[cf.][and references therein]{CuFoPi.:94,RoFeBe.:04}, that was used to create some of the data presented here.

In a second approach object line fits are in every case, but one ([\ion{O}{iii}]\,$\lambda4363$), relying primarily on an appropriate fit of the continuum. Determining the intensity of [\ion{O}{iii}]$\,\lambda4363$ additional care is necessary, and the following procedure was adopted.

The integrated sky spectrum of {\rNGCeye} (using the outermost spaxels of tile\,2) is shown in Fig.~\ref{fig:dataanalysissky}a with an additional (Gaussian) line fit of Hg$\,\lambda4358$ in the wavelength range $4320$--$4400\,${\AA}. Not only is the continuum fit a bit offset in the redmost region, but a wing of emission remains for wavelengths redwards of the mercury line. This wing is too wide to fit with any single emission line, and is likely the product of several very weak telluric lines; also compare with the adopted sky spectrum of \citet[hereafter {\rMiClWa}, see the middle panel of fig.~3 therein]{MiClWa.:91}.

In order to achieve a better fit of the continuum, and simultaneously ``remove'' the emission wing, ten additional emission lines are added (at fixed wavelengths) to fit the sky spectrum. These additional lines are at first fitted together with the mercury line in a sky spectrum of each object; the result for {\rNGCeye} is shown in Fig.~\ref{fig:dataanalysissky}b. Thereafter the intensities of these additional lines are fixed relative to the (variable) intensity of Hg$\,\lambda4358$\footnote{The literature, including for example \citet{SaAcTh.:07} for Calar Alto, does not provide any spectra of the sky with a high enough resolution and at a sufficiently high signal-to-noise to discriminate between individual features, which is why the ten wavelengths are chosen by a best fit. We used the same ten values with all objects.}. All subsequent object line fits in this wavelength range (of H$\gamma$ and [\ion{O}{iii}]$\,\lambda4363$) are then made using this set of 11 lines, in addition to the two object lines. Provided the continuum fit is appropriate, and the sky spectrum is guaranteed not to contain any flux from the object, Fig.~\ref{fig:dataanalysissky}b shows that this method is accurate in estimating the sky contribution at the wavelength of the oxygen line.

Figures~\ref{fig:dataanalysissky}c and d show two object emission line fit examples for {\rNGCeye}. In addition to the sky emission lines H$\gamma$ is seen at $\lambda\!=\!4340\,${\AA}, and [\ion{O}{iii}] at $\lambda\!=\!4363\,${\AA}. The spectrum in panel c is sampled at $r\!=\!16\arcsec$ (arc {\it e} in tile\,1, Fig.~\ref{fig:resspxmapNGC6826}), and that in panel d at $r\!=\!26\arcsec$ (arc {\it h} in tile\,2). Note that both object lines are much weaker in panel d, where a precise measurement of [\ion{O}{iii}]$\,\lambda4363$ would be impossible if the sky emission would not be given an appropriate consideration.

The sky subtraction method described here can be used with any set of object and telluric lines and spectral resolutions, for as long as the two components are resolved. Whether it is necessary to fit additional line features, as we did with the set of 11 lines, has to be decided upon on an individual basis.

\subsection{Binning of spaxels -- increasing the signal-to-noise}\label{sec:dataanalysisspxmap}
There are several ways to add individual spectra on the IFU. In regions where emission lines are strong enough it is, for each spaxel, possible to attain a surface brightness with an associated error. With such a two-dimensional signal-to-noise map it is then possible to bin the data further using, e.g., Voronoi tessellations \citep{CaCo:03}, if necessary. With weak lines, however, which have to be binned before surface brightnesses can be extracted, that approach is not possible, unless a strong line is relied upon to create the binning map. Moreover, when the signal-to-noise changes very rapidly across the IFU surface -- as it does with strong gradients -- such a binning provides only a small improvement. Here we wanted to achieve both an increase in the signal-to-noise as well as a high spatial resolution. We therefore used two approaches simultaneously.

\begin{figure*}
  \centering
  \includegraphics[height=18cm,angle=90]{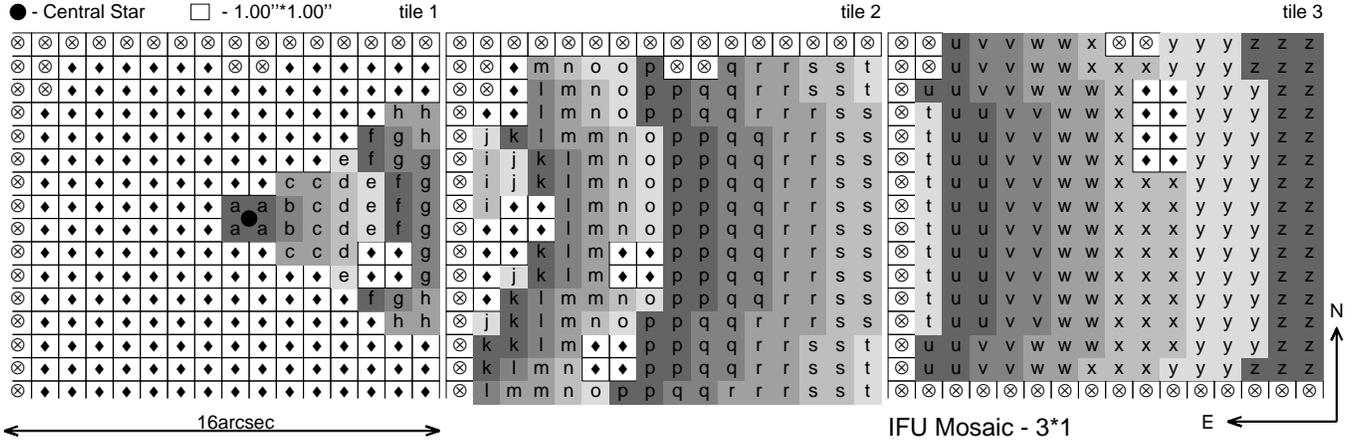}
  \caption{Binning map used in the analysis of our observations of {\rNGCblue}. Each spaxel (square) represents an area of $1\farcs0\times1\farcs0$ on the sky, and data is here present out to a distance of {36\arcsec} from the CS ($\bullet$). Each of the three tiles, in the east-west direction, has $16\times16$ elements and represents the shape of the PMAS LARR IFU; in order to emphasize that each tile is observed separately they are slightly separated from each other in the figure. Spectra of spaxels marked with the same letter are summed up to increase signal-to-noise (cf.\ Sect.~\ref{sec:dataanalysisspxmap}) -- forming concentric arcs of spaxels, which are centered on the CS. Gray shades of the spaxels were chosen to enhance the shape of the arcs. Spaxels marked with the symbol $\otimes$ are not used. The N-E most sides of spaxels in tiles 1 \& 2, together with the S-E most sides of tile 3 were all masked to not be used since they, due to the DAR correction procedure, depend on flux outside the observed area (see Sect.~\ref{sec:datareduction}). Spaxels marked with the symbol $\blacklozenge$ were not used since either one or several emission lines were removed as cosmic rays in the data reduction; note that the DAR correction procedure causes each non-wanted spaxel to result in four masked spaxels. The $\blacklozenge$ symbol is also used to mask most of tile 1, where the object structure is asymmetric. This map can be compared with the object image in Fig.~\ref{fig:obsimages}a, and also the line intensity maps in Fig.~\ref{fig:resNGC7662lmap}.}
\label{fig:resspxmapNGC7662}
\end{figure*}
\begin{figure*}
  \centering
  \includegraphics{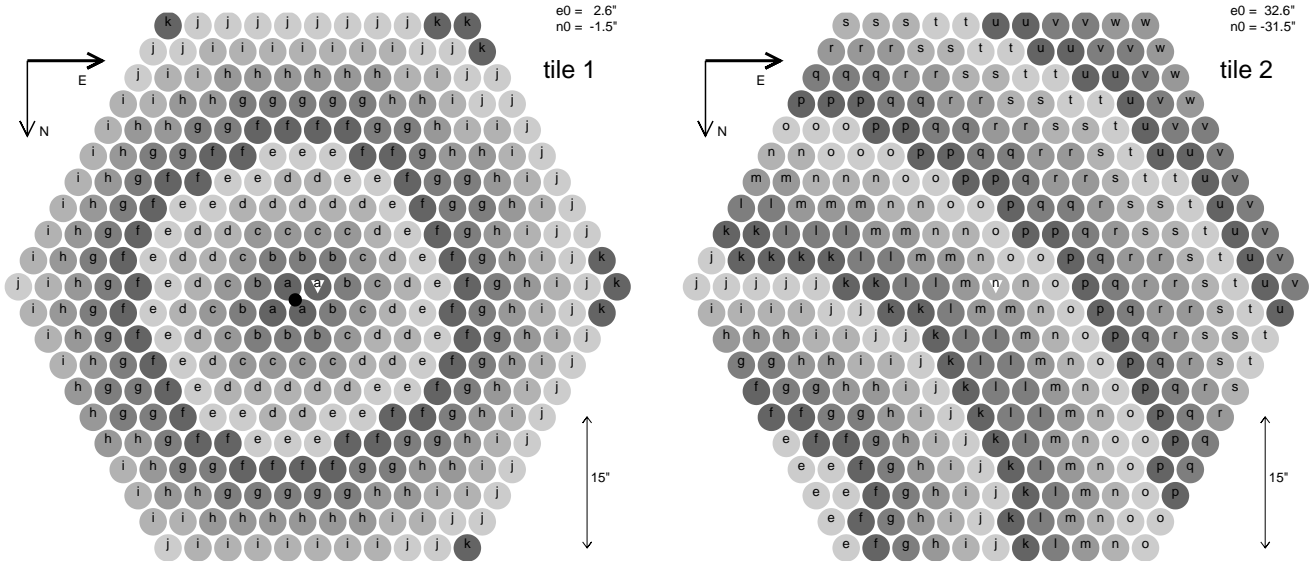}
  \caption{Binning map used in the analysis of the observations of {\rNGCeye}. Note that north is down and east to the right. In difference to the {\tLARR} IFU, which is arranged in a square structure using square-shaped lenslets, the {\tPPAK} IFU is arranged in a hexagonal structure where each fiber is circular. Each spaxel (circle) represents an area of $5\farcs64^2$ on the sky, and the outer tile (2; right) reaches a radius of 80{\arcsec} (in arc {\it w}). The coordinates indicate where the center of the respective tile (marked with a white upside down triangle) is with respect to the central star (that is indicated in the left tile with a bullet). These maps can be compared with the object image in Fig.~\ref{fig:obsimages}e, if they are flipped in both directions. For further details of the figure see the caption of Fig.~\ref{fig:resspxmapNGC7662}.}
  \label{fig:resspxmapNGC6826}
\end{figure*}

For those parts of the objects where the respective emission line is strong we created a two-dimensional map of the surface brightness. For the weaker lines in the outer parts of the objects the goal was to add up spectra which sample regions where the physical structure is (nearly) equal. In the PNe of this study all lines, but [\ion{O}{iii}]\,$\lambda\lambda4959,\,5007$, H$\alpha$, and H$\beta$ are considered weak. Note that weak emission lines in the halo will likely, at some radius, become readout-noise limited, in particular at higher spectral resolutions and when spaxels are small ($\la\!0\farcs5\times0\farcs5$). Beyond such radii these lines cannot be detected, even if spectra are binned. The halos of the objects of this study are, furthermore, all spherical to a good approximation, and our aim was therefore to extract circular (concentric) arcs of spaxels at increasing radii around a defined location of the central star (CS). As is illustrated in Sect.~\ref{sec:results}, however, real objects are typically not symmetric in the central region, where this binning consequently is less useful. A standard deviation of the intensity in each bin is calculated using the two-dimensional intensity maps, but only if an arc comprises more than three spaxels where an intensity can be measured. Such a value provides an estimate of how closely spaxels in a specific arc sample a region of similar physical conditions -- a larger value indicates a deviation from spherical symmetry in the physical structure.

A binning map illustrating the adopted configuration of spaxels for the observations of {\rNGCblue} is shown in Fig.~\ref{fig:resspxmapNGC7662}. Note that the arcs are three spaxels wide in the outermost part of the halo (to the right), in order to compensate for the lower line intensities expected there (cf.\@ Sect.~\ref{sec:resNGC7662sb}). Showing the difference in the binning setup between the {\tLARR} and {\tPPAK} observations, the two binning maps used in the analysis of the {\rNGCeye} data are shown in Fig.~\ref{fig:resspxmapNGC6826}. All arcs in both tiles are one spaxel wide, as a two-spaxel wide arc was not found to provide any significantly better signal-to noise in this case.

\subsection{Dereddening}\label{sec:dataanalysisdered}
We measured extinction coefficients using the H$\gamma$/H$\beta$ line ratio, but we also used values from the literature. In several cases we could calculate a value across the PN surface, but in the faint halos we always used a constant. In Sect.~\ref{sec:results} we compare literature and measured values for each object separately. In the dereddening procedure we used the interstellar extinction curve of \citet{FlPlTh.:94}. When measuring H$\beta$ we did not consider \ion{He}{ii} emission, since it is negligible in most cases.

\section{Results}\label{sec:results}
We made all line fits using the IFU analysis package, \textsc{ifsfit}, that we developed for this purpose. \textsc{ifsfit} is based on the \textsc{idl} routine \textsc{mpfit} of the freely available Markwardt library\footnote{http://cow.physics.wisc.edu/$^{\sim}\!$craigm/idl/fitting.html.} and was designed to work with most IFUs. Gaussian curves were used to fit emission lines, and polynomials to fit continua.

In a first step we made a line fit for each line and spaxel on every data set in order to get a surface brightness and an associated error. The outcome form two-dimensional spatial maps of surface brightnesses. Not all line fits are good, and bad fits and inaccurate fits are marked as such after a visual inspection of each individual line fit. A bad fit typically occurs when a line disappears in noise, or if an unmasked line was removed as a cosmic ray by accident during the data reduction (see Sect.~\ref{sec:datareduction}). Inaccurate fits may occur if the continuum level is badly estimated, or if the actual line width in a particular spaxel is larger or smaller than the pre-defined and allowed deviation of $\pm15$\%, from the theoretically determined full width at half maximum (of the grating in the used wavelength range).

In a second step we used a pre-defined map of radially binned spaxels (see Sect.~\ref{sec:dataanalysisspxmap}), where we averaged the contributing spectra before fitting the emission lines. The details of the fitting procedure are the same as in the first step. The outcome consists of a surface brightness and an error for each (radial) position of the respective binned arc and emission line. Surface brightness errors are typically small. When we draw error bars in plots of the radial surface brightnesses we instead use the spaxel-to-spaxel standard deviation (Sect.~\ref{sec:dataanalysisspxmap}).

Concerning physical quantities an electron temperature (\Te) was in every case calculated using the expression of \citet[see eq.~5.4 therein]{Os:06} for [\ion{O}{iii}], assuming a negligible dependence on the electron density (\ne). An electron density was derived using two methods. Where available the forbidden line ratios [\ion{Ar}{iv}]\,$\lambda\lambda4711/4740$, [\ion{O}{ii}]$\,\lambda\lambda3726/3729$, and [\ion{S}{ii}]\,$\lambda\lambda6717/6730$ were used to calculate $\nee{\text{\ion{Ar}{iv}}}$, $\nee{\text{\ion{O}{ii}}}$, and $\nee{\text{\ion{S}{ii}}}$ -- using the \textsc{iraf nebular} package of \citet{ShDu:95}. We calculated maximum error limits of the density using the minimum and maximum estimates of the respective ratio.

Since the (electron) density in the halo in general is so low that forbidden emission line ratios always are outside density sensitive ranges (\rSK) another method must be used there. Assuming spherical symmetry, full ionization, and a distance it is possible to derive an emissivity ($j$) using an (inverse) Abel transform on the radial surface brightness structure of a recombination line (such as H$\beta$). The electron density is proportional to the emissivity as $\ne\!\propto\!j^{1/2}$. This method is first used by \citet[hereafter {\rPlSo}]{PlSo:90} on {\rNGCeye} in order to calculate an electron density structure. The electron density can thereafter be converted into a mass density and a mass loss rate (see Sect.~\ref{sec:discmassloss}). Since our density estimates are based on the observed intensity structure, and not on the true (deconvolved) intensity, they are partly affected by scattered light, which is the strongest in the inner halo (cf.\@ Sect.~\ref{sec:dataanalysisstray}).

We present the results for each object separately. In consecutive order they are surface brightnesses, electron temperatures, and electron densities.

\subsection{{\roNGCblue} -- PN\,G106.5-17.6}\label{sec:resNGC7662}
The spaxel binning map that we used with the data of this object to calculate a radial intensity structure is shown in Fig.~\ref{fig:resspxmapNGC7662}. In the central region {\rNGCblue} is asymmetric, which is why most of the spaxels there were masked, in order to achieve a smoothly varying surface brightness across the inner-middle tile (1--2) boundary. A recent detailed discussion of structural properties of {\rNGCblue} is given by \citet{GuJaCh:04}. Moreover, a distance is determined by \citet[$d\!=\!1.11$--$1.24\,$kpc, who includes a literature summary]{Zh:95}, \citet[$d\!=\!0.79\!\pm\!0.75\,$kpc]{HaTe:96}, and {\rCSSP} ($d\!=\!2.2\,$kpc, from G\'orny priv.\@ comm.). {\rMiClWa} use $d\!=\!1.5\,$kpc. We used $d\!=\!1.7\,$kpc (see Sect.~\ref{sec:discmassloss}). A value of the effective temperature of the CS is given by G\'orny (priv.\@ comm.), $\teff(\text{Zanstra})\!=\!100\,000\,$K.

\begin{figure}
\includegraphics{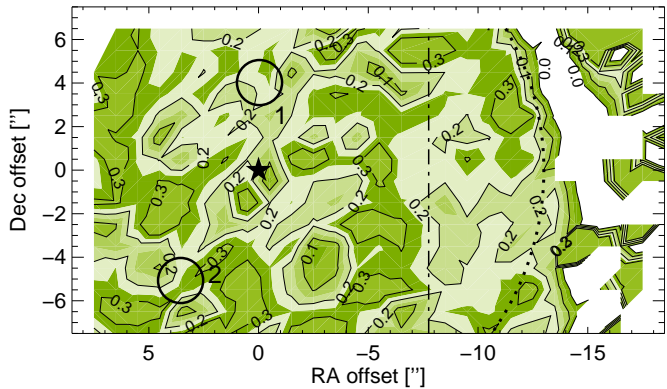}
\caption{A map of the calculated extinction coefficient $c$ of {\rNGCblue} for tiles 1 and 2. The CS position is marked with a filled star; the two numbered circles indicate regions observed by \citet{Ba:86}. On the east and north sides the ``missing'' column and row of values were removed in the DAR correction procedure. The vertical dash-dotted line marks the tile boundary. For further details see Sect.~\ref{sec:resNGC7662}.}
\label{fig:rescextmap}
\end{figure}

Numerous references give one (or a few) scalar extinction coefficient for this extended object, see, e.g., \citet[$c\!=\!0.42$]{O:64}, \citet[$c\!=\!0.16\!\pm\!0.12$]{PeTo:71}, \citet[$c\!=\!0.23$]{HaSeAd.:82}, \citet[$c\!=\!0.17$, hereafter {\rKa}]{Ka:86}, {\rMiClWa} ($c\!=\!0.23$), \citet[$c\!=\!0.16\!\pm\!0.04$]{LaPo:96}, and \citet[$c\!=\!0.10$]{HyAl:97}. We first calculated a map for the central region of the nebula, see Fig.~\ref{fig:rescextmap}. Two out of six positions where \citet[$c_{(1)}\!=\!0.26$, $c_{(2)}\!=\!0.17$]{Ba:86} measures the extinction coefficient are also shown in the map, indicating a fair agreement. A vague outline of the rim is also visible, where the extinction is slightly higher than elsewhere. Comparing this map with that presented in \citet[fig.~3a, that uses the same data, but does not correct for DAR]{SaScRo.:07}, it is seen that there is less structure in this case. The average of the values in the central tile is $c\!=\!0.24\pm0.09$, which is close to the value measured by {\rMiClWa}. In the outer parts of the nebula, where an extinction coefficient cannot be determined (as in the westmost side of Fig.~\ref{fig:rescextmap}), we always use this constant value.

\begin{figure*}
\sidecaption
\includegraphics[width=12cm]{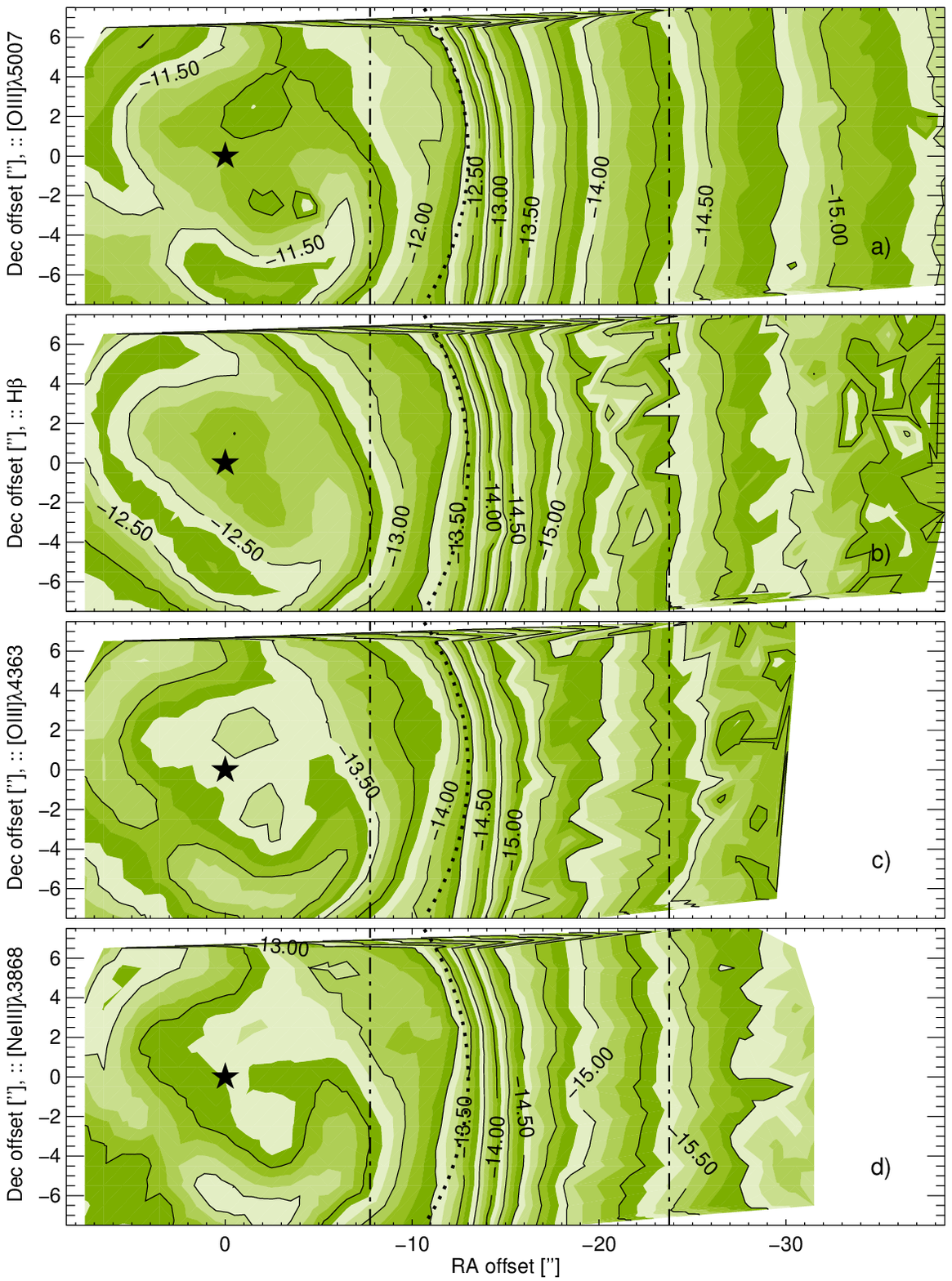}
\caption{Surface brightness maps of a mosaic of the three observed tiles of {\rNGCblue} for four emission lines. Intensities are logarithmic and the unit is {\sb}. The vertical dash-dotted lines mark the separation between the three tiles. The position of the central star is marked with the symbol $\bigstar$. Colors repeatedly show additional contour lines; the separation between two color contours corresponds to a $0.083\,\mbox{dex}\!=\!21$\% change in surface brightness. Dense lines in the northmost row of spaxels, in the middle tile, are created by the contour drawing routine, and do not show a real gradient. From the top the panels show four increasingly blue lines: {\bf a)} [\ion{O}{iii}]$\,\lambda5007$, {\bf b)} H$\beta$, {\bf c)} [\ion{O}{iii}]$\,\lambda4363$, and {\bf d)} [\ion{Ne}{iii}]$\,\lambda3869$. Corresponding signal-to-noise maps for [\ion{O}{iii}]$\,\lambda5007$ and [\ion{O}{iii}]$\,\lambda4363$ are shown in Fig.~\ref{fig:resNGC7662lmape}. The dotted line in the middle tile indicates the location of the shell-halo transition ($r_{\text{shell}}\!=\!13\arcsec$). For further details see Sect.~\ref{sec:resNGC7662sb}.}
\label{fig:resNGC7662lmap}
\end{figure*}

\begin{figure*}
\includegraphics[width=18.3cm]{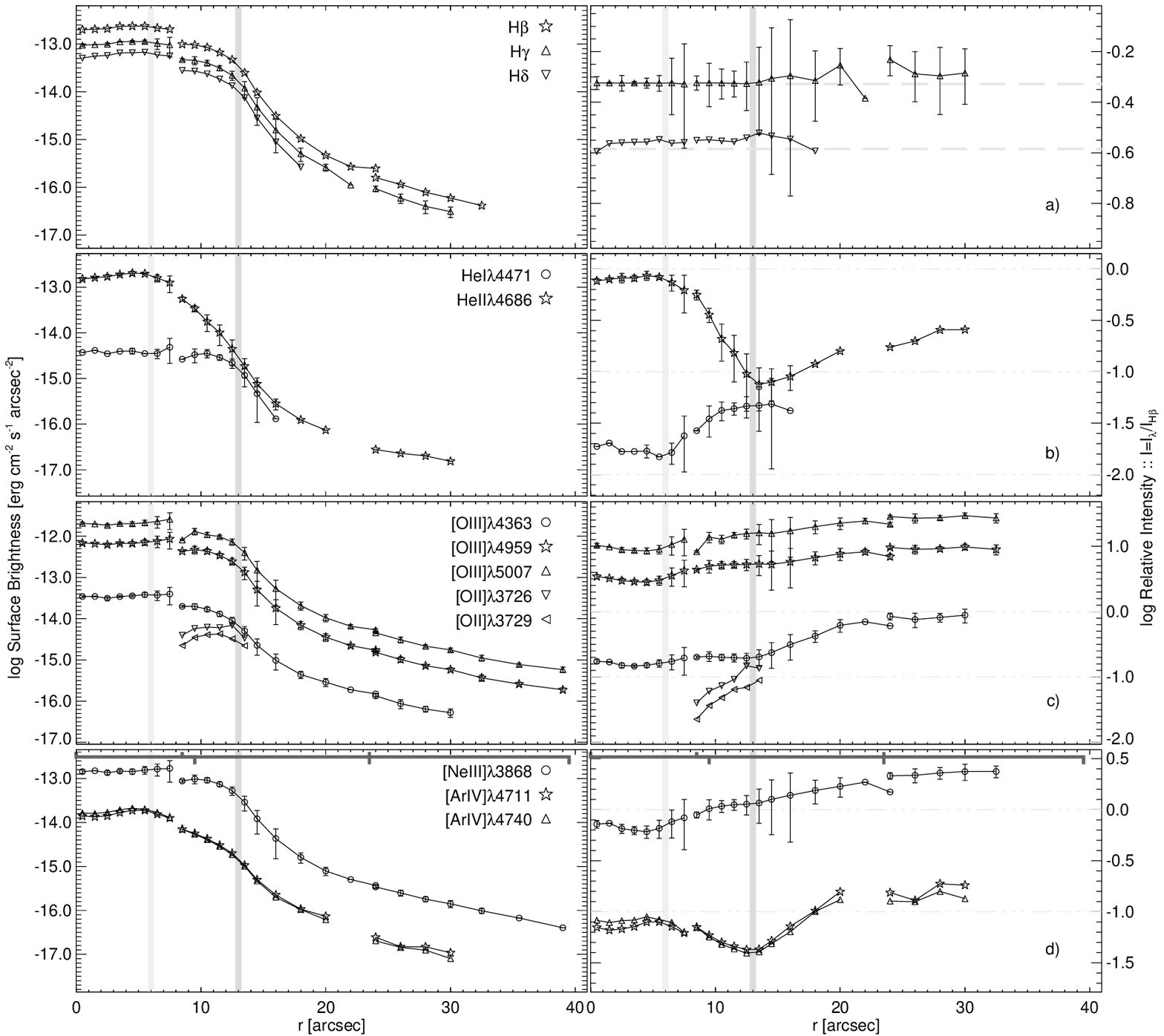}
\caption{Radial surface brightness structures of {\rNGCblue}, cf.\ Sect.~\ref{sec:resNGC7662}. All ordinates are logarithmic. Intensities of each ion are indicated using the same symbol in both left and right hand side panels. Panels on the right hand side show the log of each intensity ratio relative to H$\beta$. From the top the panels show 13 emission line intensities of five elements: {\bf a)} hydrogen, {\bf b)} helium, {\bf c)} oxygen, and {\bf d)} neon and argon. Error bars due to Poisson and readout noise were omitted since they in every case are smaller than the symbol sizes. Instead error bars indicate the spaxel-to-spaxel standard deviation -- for clarity only for a few lines (cf.\ Sect.~\ref{sec:results}). The radial positions of the three tiles are indicated in the upper part of panel {\bf d}. For each tile the inner edge is marked by an upwards pointing marker and the outer edge by a downwards pointing marker. Note the overlap in the data between tiles 1 \& 2 and 2 \& 3 (see Fig.~\ref{fig:obsimages}a for the actual positions of the tiles on the sky). Vertical gray bars indicate the location of the rim at $r_{\text{rim}}\!=\!6\arcsec$, and the shell-halo transition at $r_{\text{shell}}\!=\!13\arcsec$. The horizontal dashed gray lines in the right {\bf a} panel indicate line ratios for Case B recombination theory at $\Te\!\approx\!10\,000$\,K. Horizontal dash-triple-dotted gray lines in the right hand side panels are guides, that indicate the constant ratios $0.01$, $0.1$, and $1.0$. Data is missing for some of the weaker lines at $r\approx$20{\arcsec}, since the exposure time of tile 2 (200\,s) was not long enough to measure them accurately. The offset between data of tiles 1 and 2 at $r\!\approx\!9\arcsec$ (and tiles 2 and 3 at $r\!\approx\!24\arcsec$), is due to deviations from symmetry.}
\label{fig:ressbNGC7662}
\end{figure*}

\subsubsection{Surface brightnesses}\label{sec:resNGC7662sb}
Two-dimensional surface brightness maps of four selected emission lines are shown in Fig.~\ref{fig:resNGC7662lmap}. The (asymmetric) elliptical rim is clearly seen (in the central PN, i.e. the leftmost tile) along with a rapidly radially outwards decreasing intensity. Each panel shows contours of the respective logarithmic intensity, revealing the geometrical shape of the nebula at different intensity levels. In the two outer tiles intensity contours (which are circular to first order) are present almost all the way to the west-most (right) boundary in the strong oxygen line [\ion{O}{iii}]$\,\lambda5007$ (Fig.~\ref{fig:resNGC7662lmap}a). H$\beta$ (Fig.~\ref{fig:resNGC7662lmap}b) shows larger fluctuations in the halo already in the middle tile. Also note that this recombination line is structured differently on the rim compared to the other three lines; the rim is closed on the southeast side of the CS. [\ion{O}{iii}]$\,\lambda4363$ and [\ion{Ne}{iii}]$\,\lambda3869$ (Figs.~\ref{fig:resNGC7662lmap}c and d), finally, are both weak and were not detected either in the outer part of the middle tile or in the outermost tile.

Corresponding signal-to-noise maps are shown in Fig.~\ref{fig:resNGC7662lmape} for the two oxygen lines. The discontinuous contours, that are seen at tile boundaries in Fig.~\ref{fig:resNGC7662lmape}, indicate a changing signal-to-noise with exposure time. Such a discontinuity is not visible between the outer two tiles of the [\ion{O}{iii}]\,$\lambda4363$ line that is shown in Fig.~\ref{fig:resNGC7662lmape}b. In this case no line could be measured in the outer parts of the middle tile. Additionally, note the intensity fluctuations at lower signal-to-noise values (S/N\,$\la\!10$) as seen in the rightmost tile (3) of Fig.~\ref{fig:resNGC7662lmap}b and Fig.~\ref{fig:resNGC7662lmape}b. Similar intensity maps were created for all lines and objects, but are not all presented due to limitations of space.

In Fig.~\ref{fig:ressbNGC7662} we present radial surface brightness structures that were derived from the binned arcs for 13 emission lines. Note that we also could measure the unresolved oxygen line doublet [\ion{O}{ii}]$\,\lambda\lambda3726,\,3729$ in the inner region of the central tile. Most lines drop in intensity by a factor of $3000$ and more between the central parts and the outermost parts of the observed halo. The argon and helium lines are not present in the region around $r\!\approx\!21\arcsec$, since the exposure time used with tile 2 (200\,s) did not allow a detection in these lines. Due to the asymmetric shape of the central nebula, line intensities in adjacent spaxels of tiles 1 and 2 differ by as much as 50 per cent, explaining the rather large step in intensities at $r\!\simeq8\!\arcsec$. Likewise, signs of asymmetry are also visible for intensities at $r\!\simeq\!24\arcsec$ (arc \emph{t}) in tiles 2 and 3; this arc comprises five and ten spaxels, respectively. The size of the error bars in the right hand side panels show that the departure from spherical symmetry in separate bins is the largest in all lines for $6\arcsec\!\la\!r\!\la\!8\arcsec$ and $12\arcsec\!\la\!r\!\la\!16\arcsec$.

Comparing the PN proper and the halo further the relative intensities (right hand side panels in Fig.~\ref{fig:ressbNGC7662}) show an increasing trend in the halo outwards from the shell-halo transition, i.e.\ for $r\!\ga\!13\arcsec$, in all (but the hydrogen) lines. The intensity ratio H$\gamma$/H$\beta$ (H$\delta/\text{H}\beta$) is, where a two-dimensional extinction coefficient map could be used, constant to a good approximation. In Case B ionization the, relatively temperature insensitive, ratio is 0.47 (0.26) for $\Te\!=\!10\,000$\,K \citep[see, e.g., table~4.4 in][]{Os:06}. While the H$\gamma/$H$\beta$-ratio is close to the theoretical value the H$\delta/$H$\beta$-ratio is off by about $15$ per cent out to the shell-halo transition. This difference indicates a systematic error in the surface brightness of H$\delta$, the agreement between theory and observations of this ratio is better with the other objects. Emission lines are in every case, but H$\delta$ and \ion{He}{i}\,$\lambda4471$, present at intensities below $1\!\times\!10^{-16}\,\sb$. The sky continuum intensity is about $1\!\times\!10^{-17}\,\sb$, which is the sensitivity limit achieved with the adopted instrumental setup and exposure times\footnote{For the bluest measured emission line, [\ion{Ne}{iii}]\,$\lambda3869$, the sky continuum intensity is about $2\!\times\!10^{-17}\,\sb$.}. The amount of scattered light is small, this is confirmed when comparing surface brightnesses with the radial intensity level in fig.~21 of {\rCSSP} ([\ion{O}{iii}] and [\ion{O}{iii}]$-$\,diffuse light, where it was necessary to subtract scattered light present in JKT data).

\begin{figure}
\includegraphics{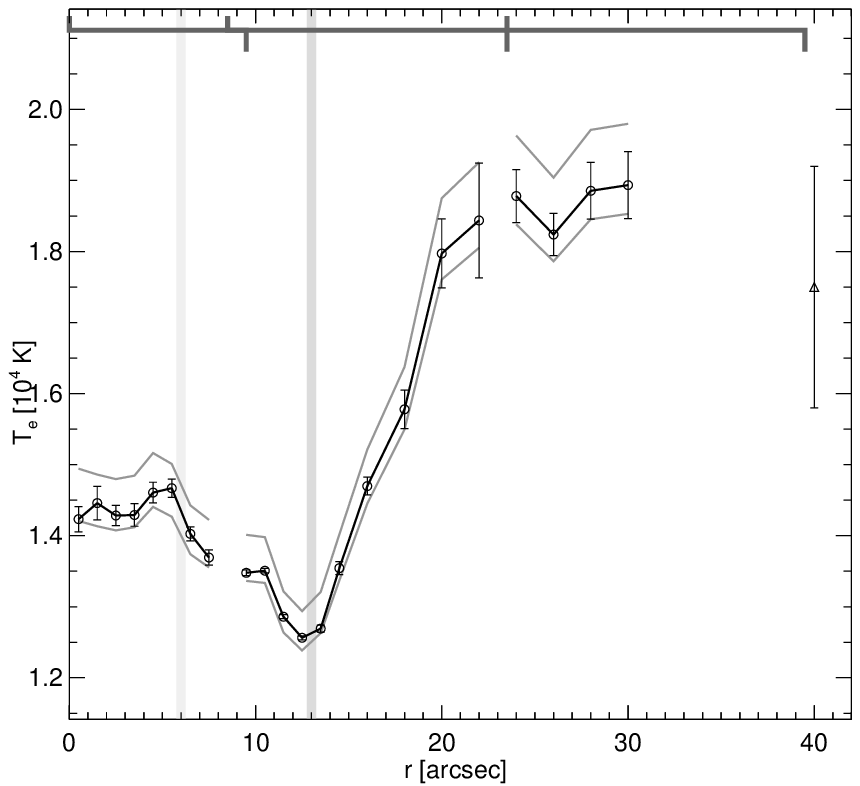}
\caption{Radial temperature structure of {\rNGCblue}. In addition the halo measurement of {\rMiClWa} is indicated with a triangle ($\triangle$). The inner gray vertical line indicates the location of the rim, and the outer line the shell-halo transition. The upper (lower) gray line shows {\Te} calculated with the extinction coefficient set to $c\!=\!0.4$ ($0.1$). This radial structure can be compared with the two-dimensional temperature map of the inner parts of the nebula in Fig.~\ref{fig:resNGC7662Temap}. For further details see Sect.~\ref{sec:resNGC7662T}.}
\label{fig:resNGC7662Te}
\end{figure}

\subsubsection{Electron temperature}\label{sec:resNGC7662T}
We could measure [\ion{O}{iii}]$\,\lambda4363$ out to a distance of $r\!=\!30\arcsec$ from the CS. In the calculation of the temperature we used all three diagnostic [\ion{O}{iii}] lines. The radial structure is presented in Fig.~\ref{fig:resNGC7662Te}. Additionally, we draw two gray lines in the figure, that were derived using a constant extinction coefficient; $c\!=\!0.1$ ($0.4$) was used for the lower (upper) line. These two lines demonstrate how sensitive the measured temperature is to the use of the extinction coefficient. The difference in the halo (for $r\!\ga\!13\arcsec$), where the value of the actual extinction coefficient is more uncertain, is about 500\,K between the black line and the lower gray line.

From the shell-halo transition, and outwards in the halo, the temperature increases from $\Te\!=\!12\,500\pm100$\,K, at $r\!=\!13\arcsec$, to $\Te\!=\!18\,500\pm1000\,$K, for $r\ga22\arcsec$. Moving inwards, instead, the temperature also increases, from $\Te\!=\!12\,500\,$K, at $r\!=\!13\arcsec$, to about $\Te\!=\!14\,700\pm300$\,K at the location of the rim; due to the asymmetric inner region this increase is not smooth (see below).

The halo measurement of {\rMiClWa}, $\Te\!=\!17\,500\pm1700$\,K, is also indicated in Fig.~\ref{fig:resNGC7662Te}. It was made in about the same region of the halo that is covered here. The agreement with our value at $r\!=\!30\arcsec$ is fairly good. Additionally, {\rMiClWa} quote an average core temperature of $\Te\!=\!13\,100\pm500$\,K (where the measurement of [\ion{O}{iii}]$\,\lambda4363$ is taken from \citealt{PeTo:71}). \citet{HyAl:97} derive $\Te\!\simeq\!12\,500$\,K (using $c\!=\!0.10$ and assuming $\ne\!=\!8000\,\cbm$) in about the same region as position 2 of \citet{Ba:86}. Further estimates are presented by, e.g., {\rKa}, \citet{PeLuTo:95} and \citet{ZhLiWe.:04}. The values are in general about 1000\,K lower than what we found here in the central nebula.

\begin{figure*}
\sidecaption
\includegraphics[width=12cm]{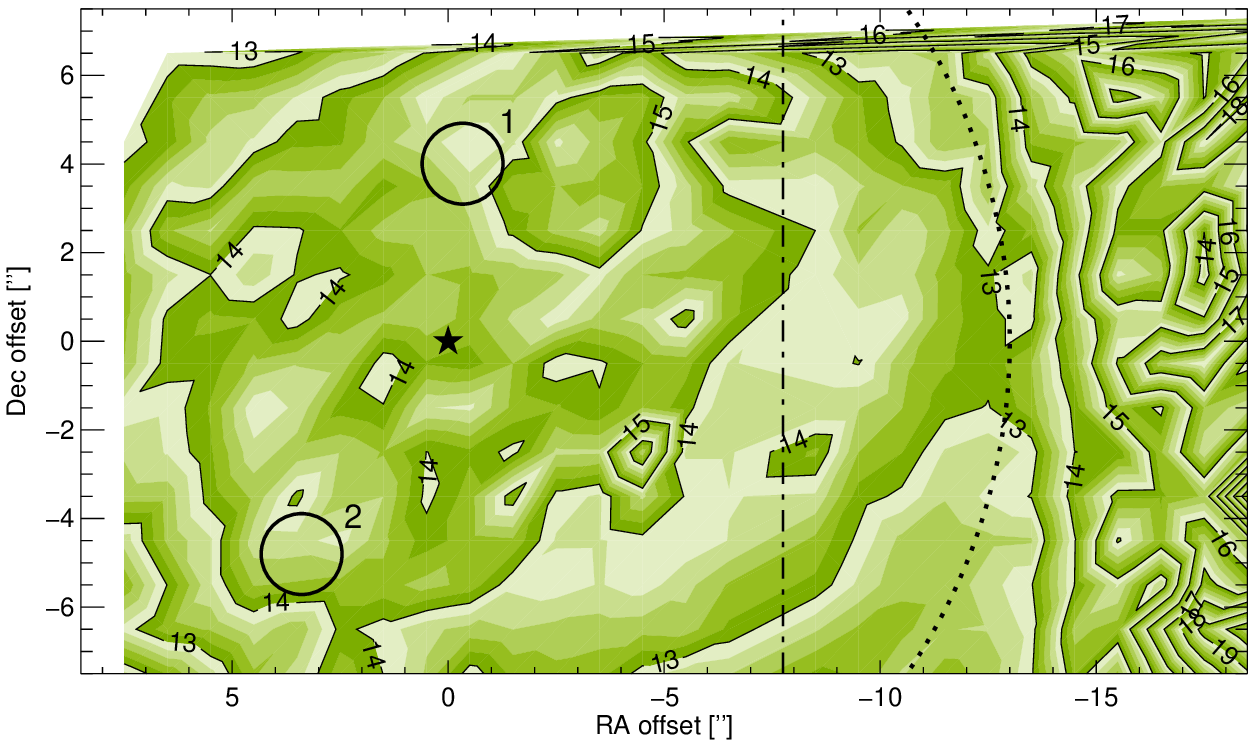}
\caption{Electron temperature map of the central regions of {\rNGCblue}, shown in the unit of $10^3\,$K. Colors repeatedly indicate additional contours at a 200\,K interval. The vertical dash-dotted line marks the separation between tiles 1 and 2. The position of the CS is marked with a filled star. Temperatures are measured in the interval $11\,800$--$16\,000$\,K. The location of the shell-halo transition is marked with a dotted arc, $r_{\text{shell}}\!=\!13\arcsec$. The lowest values are found in the outer shell and the highest values in regions inside the rim, compare with the surface brightness maps in Figs.~\ref{fig:resNGC7662lmap}a \& c. The two numbered circles indicate regions also observed by \citet{Ba:86}. For further details see Sect.~\ref{sec:resNGC7662T}.}
\label{fig:resNGC7662Temap}
\end{figure*}

In comparison to the binned radial temperature structure we show a two-dimensional temperature map of the central regions in Fig.~\ref{fig:resNGC7662Temap}. A comparison is best made with the central parts of the map shown in fig.~4 of \citet[that is derived using an electronographic method]{ReWo:82}. The agreement is good -- the asymmetric region extending to the west is clearly seen in both maps. Absolute values are also in better agreement than with the other references. \citet{Ba:86}, moreover, derives a temperature at five different positions across the central nebula, where two are covered in Fig.~\ref{fig:resNGC7662Temap} (positions 1 [4{\arcsec}\,N] and 2 [3\farcs7\,E, 5{\arcsec}\,S]). Both values, $13\,800\pm500$\,K and $13\,100\pm500$\,K, respectively, are again about 1000\,K lower than what is shown in Fig.~\ref{fig:resNGC7662Temap}. In the halo the low signal-to-noise in each spaxel of the weak [\ion{O}{iii}]\,$\lambda4363$ prevents the calculation of a map, instead it is necessary to bin the spaxels, as we did for Fig.~\ref{fig:resNGC7662Te}.

\subsubsection{Electron density}\label{sec:resNGC7662ne}
At first we measured the electron density at a few radial locations using the [\ion{Ar}{iv}]\,$\lambda4711/4740$ and [\ion{O}{ii}]\,$\lambda3726/3729$ line ratios. Using the oxygen doublet we derived ratios for $r\!=\!8\farcs5,\,9\farcs5$, and $10\farcs5$ to find $\nee{\text{\ion{O}{ii}}}\!=\!1800^{+740}_{-580},\,1540^{+320}_{-280}$, and $1260^{+160}_{-150}\,\cbm$. Due to the low resolution it is difficult to deblend [\ion{Ar}{iv}]$\,\lambda4711$ and the, in this case, very weak \ion{He}{i}$\,\lambda4713$. The resulting values for $r\!=\!8\farcs5,\,9\farcs5,\,10\farcs5,\,11\farcs5$, and $12\farcs5$ were, $\nee{\text{\ion{Ar}{iv}}}\!=\!4540\pm230,\,4110\pm200,\,3950\pm170,\,3760\pm160$, and $3460\pm180\,\cbm$. Our values are in good agreement with \citet{LaPo:96}, who measure $\nee{\mbox{\ion{S}{ii}}}\!=\!4000\,\cbm$ in the shell. Other authors report on values measured in regions not considered here, e.g.: \citet[$\nee{\text{\ion{Ar}{iv}}}(1,2)\!=\!6300,\,7000\,\cbm$ and $\nee{\text{\ion{Cl}{iii}}}(1,2)\!=\!3400,\,1500\,\cbm$]{Ba:86}.

\begin{figure}
\includegraphics{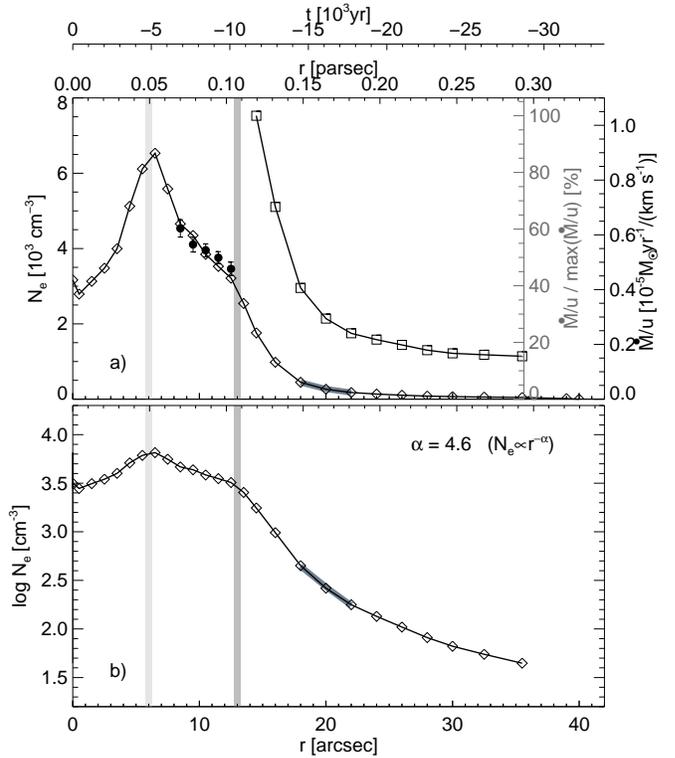}
\caption{Radial density structure of {\rNGCblue} (diamonds, $\diamondsuit$) calculated using an Abel transform. The structure is shown with a logarithmic ordinate in panel {\bf b} (bottom). In panel {\bf a} (top) our mass loss rate estimate $\dot{M}/u$ is also shown for a part of the radial extent (using squares and the two right axes, cf.\@ Sect.~\ref{sec:discmassloss}). The gray axis shows the mass loss rate normalized with the maximum plotted value ($\dot{M}_{\text{max}}$), in per cent. The temporal axis above the plot shows the age of the structure radially, assuming a constant outflow speed of $u\!=\!10\,\kms$. The thick gray line shows a fit of the density to the power law $\ne\!\propto\!r^{-\alpha}$ for the region $17\arcsec\!\le\!r\!\le\!22\arcsec$. Bullets mark individual measurements of the electron density $\nee{\text{\ion{Ar}{iv}}}$, along with maximum error estimates. The inner gray vertical line indicates the location of the rim, and the outer line the shell-halo transition. For further details see Sect.~\ref{sec:resNGC7662ne}.}
\label{fig:resNGC7662ne}
\end{figure}

On the other hand we calculated a density structure using the inverse Abel transform with H$\beta$. The resulting electron density structure is shown in Fig.~\ref{fig:resNGC7662ne}, together with the four measured values of the argon line ratio, against which we scaled the density structure. In drawing the upper axis we used the distance $d\!=\!1.7$\,kpc. Note that the density is less accurate for the central and outer regions ($10\arcsec\!\ga\!r\!\ga\!30\arcsec$) due to the asymmetric center and an H$\beta$ structure, which is truncated at $r\!=\!35\arcsec$ (since we did not cover the full radial extent of the halo).

\subsection{{\roIC} -- PN\,G123.6+34.5}\label{sec:resIC3568}
{\rIC} is a close to round nebula with a measured size of $r_{\text{shell}}\!=\!9\arcsec$ \citep[][who used observational data of \citealt{CaDa:60}]{CoDaOd:61}, or $r_{\text{shell}}\!=\!6\arcsec$ \citep[see][and references therein]{ZhKw:93}. Distance determinations, including literature summaries, are made by, e.g., \citet[$d\!=\!2.68$--$3.12\,$kpc]{Zh:95} and \citet[$d\!=\!1.08$--$3.91\,$kpc]{Ph:05}. {\rCSSP} quotes a Shklovsky distance of $d\!=\!4.5\,$kpc, we used $d\!=\!4.3\,$kpc (see Sect.~\ref{sec:discmassloss}). The effective temperature of the CS is determined to be $\teff\!=\!50\,000\,$K \citep{HaFe:83}, $\teff\!=\!51\,300\,$K \citep{PrAcKo.:91}, and $\teff(\text{Zanstra})\!=\!55\,000\,$K (G\'orny priv.\@ comm.).

Values of the extinction coefficient are given by, e.g., \citet[$c\!=\!0.25$]{CoDaOd:61}, \citet[$c\!=\!0.10\pm0.12$]{PeTo:71}, \citet[$c\!=\!0.23\!\pm\!0.05$]{SiKa:76}, \citet[$c\!=\!0.29\!\pm\!0.04$]{PoWeWu.:77}, \citet[$c\!=\!0.252\!\pm\!0.032$]{Ba:78}, \citet[$c\!=\!0.25$]{HaFe:83}, {\rKa} ($c\!=\!0.19$), \citet[$c\!=\!0.4$]{TyAcSt.:92}, \citet[$c\!=\!0.45$, measured $4\arcsec\,$N of the CS, hereafter {\rKwHe}]{KwHe:98} and \citet[$c\!=\!0.26$]{LiLiLu.:04}, who all report on values which are measured at or close to the nebular center. Our measurements in the central tile revealed a varying coefficient, that is $c\!\ga\!0.4$ $4\arcsec$ west of the CS, $c\!\la\!0.2$ $4\arcsec$ north of the CS, and negative values are found east of the CS. Instead of using a map we chose to set $c\!=\!0.2$ across the entire surface of the nebula. Also \citet{PhReWo:79} present evidence for a differential extinction.

The two binning maps that we used with the data of this object are illustrated in Fig.~\ref{fig:resspxmapIC3568}. These maps are identical with the exception of the DAR mask, which is different for wavelengths blue- and redwards of the selected reference wavelength ($\lambda_{\text{ref}}\!=\!5100\,$\AA; cf.\@ Sect.~\ref{sec:datareduction}). Data are here present out to a distance of $r\!=\!25\arcsec$ from the CS.

\subsubsection{Surface brightnesses}\label{sec:resIC3568sb}
\begin{figure*}
\includegraphics{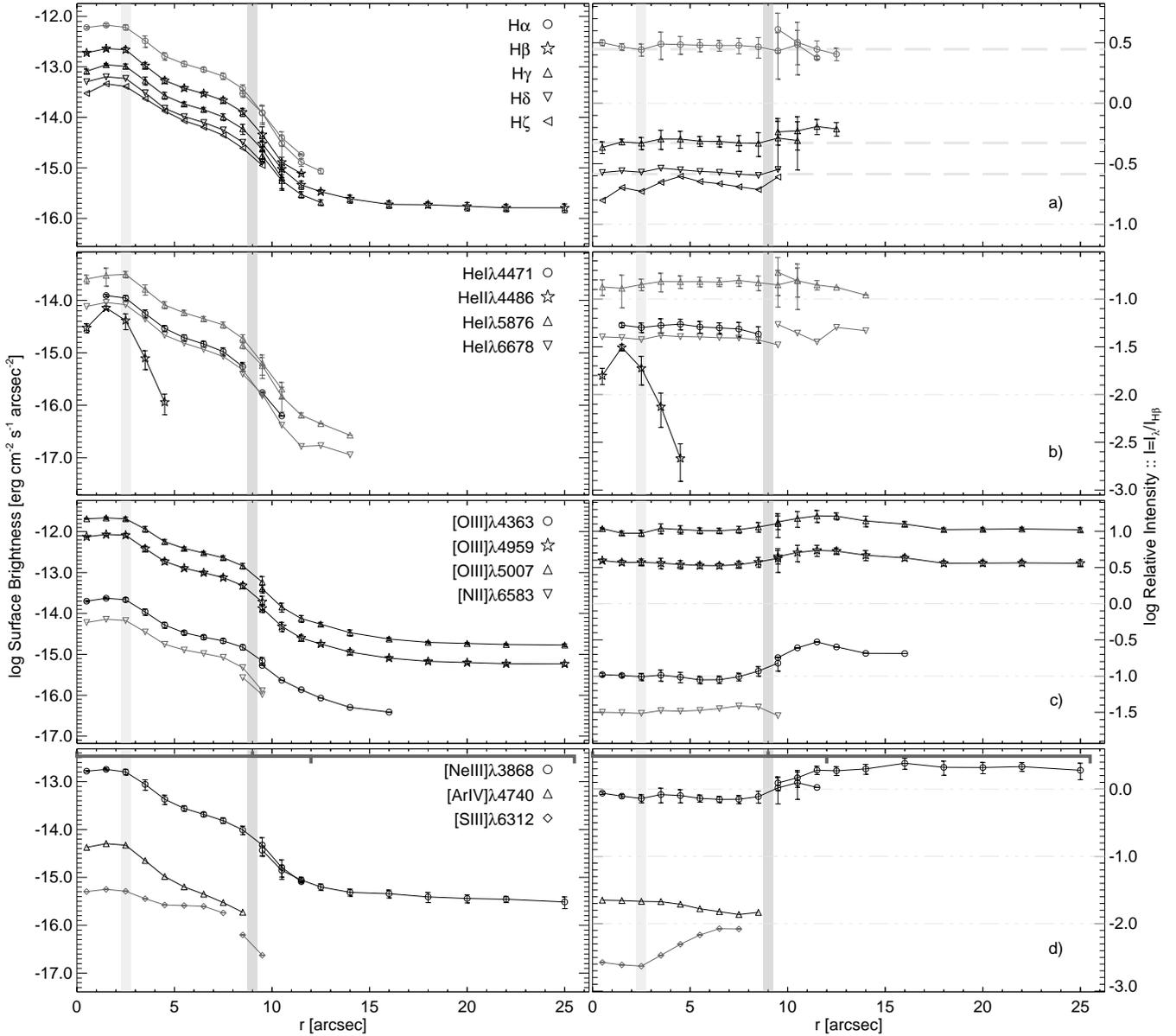}
\caption{Radial surface brightness structures of {\rIC}, cf.\@ Sect.~\ref{sec:resIC3568sb}. From the top the panels show 16 emission lines of seven elements: {\bf a)} hydrogen, {\bf b)} helium, {\bf c)} oxygen and nitrogen, and {\bf d)} neon, argon, and sulfur. For further details of figure properties see Fig.~\ref{fig:ressbNGC7662}. Vertical gray bars indicate the location of the rim at $r_{\text{rim}}\!\simeq\!2\farcs5$, and the shell-halo transition at $r_{\text{shell}}\!\simeq\!9\arcsec$. Emission lines with wavelengths longer than $\lambda_{\text{ref}}\!=\!5100\,${\AA} are drawn with gray lines, remaining lines are drawn in black. Also see Sect.~\ref{sec:resIC3568}.}
\label{fig:ressbIC3568}
\end{figure*}

Radial surface brightness structures, which were derived using the binned arcs, are presented in Fig.~\ref{fig:ressbIC3568}. From the spatial structures of the lines the location of the rim and the shell-halo transition were determined to be $r_{\text{rim}}\!\simeq\!2\farcs5$ and $r_{\text{shell}}\!\simeq\!9\farcs0$, respectively. Physical properties differ in the spaxels of those arcs, that are present in both tiles 1 and 2, causing a step of $\la\!0.2\,$dex in the surface brightness structure for $9\farcs0\!\la\!r\!\la\!12\arcsec$. In addition to the presented lines [\ion{N}{ii}]$\,\lambda6548$ and [\ion{Ar}{iv}]$\,\lambda4711$ were also detected, but could not be accurately determined. The nitrogen line is very weak compared to H$\alpha$, and the argon line cannot be deblended from \ion{He}{i}$\,\lambda4713$ due to the low resolution.

H$\beta$, the two oxygen lines [\ion{O}{iii}]$\,\lambda\lambda4959,\,5007$ and neon [\ion{Ne}{iii}]$\,\lambda3868$, are all present in the entire region covered. The surface brightnesses in the outermost arc in the halo are about $0.6$--$2\times10^{-3}$ times as strong as on the rim. Also note that the shell, compared to the rim, is weaker than for the other objects; the corresponding intensity ratios are about 25, instead of about 5--8. Two out of three of the relevant halo selection criteria of {\rCSSP} are met. The first one is not, since we did not cover the full radial extent of the assumed halo. Since the [\ion{O}{iii}] emission is too strong in the outer regions, to be caused by scattered light, we conclude, in contrast to the result of {\rCSSP}, that {\rIC} has a halo.

\subsubsection{Electron temperature}\label{sec:resIC3568T}
\begin{figure}
\includegraphics[bb=172 342 421 591,angle=90]{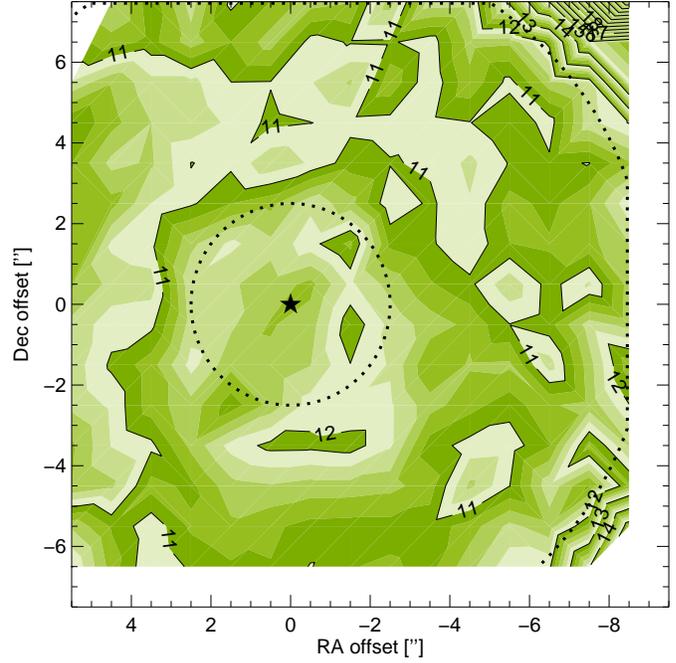}
\caption{Electron temperature map of the central regions of {\rIC}, in the unit of $10^3\,$K, cf.\@ Sect.~\ref{sec:resIC3568T}. The location of the rim is marked with the inner dotted circle, and the shell-halo transition with the outer circle. For further details of the figure see Fig.~\ref{fig:resNGC7662Temap}.}
\label{fig:resIC3568Temap}
\end{figure}

We calculated a temperature using all three diagnostic [\ion{O}{iii}] lines. A two-dimensional map of the central region is shown in Fig.~\ref{fig:resIC3568Temap}, revealing a slightly asymmetric structure. We measured temperatures in the interval $10\,400\!\la\!\Te\!\la\!12\,000$\,K. A similar temperature map is presented by \citet[$10\,800\!\le\!\Te\!\le\!13\,200\,$K]{ReWo:82}. A comparison between the two maps shows few similarities, although there is a match in the existence of a region with lower temperatures east of the CS.

\begin{figure}
\includegraphics{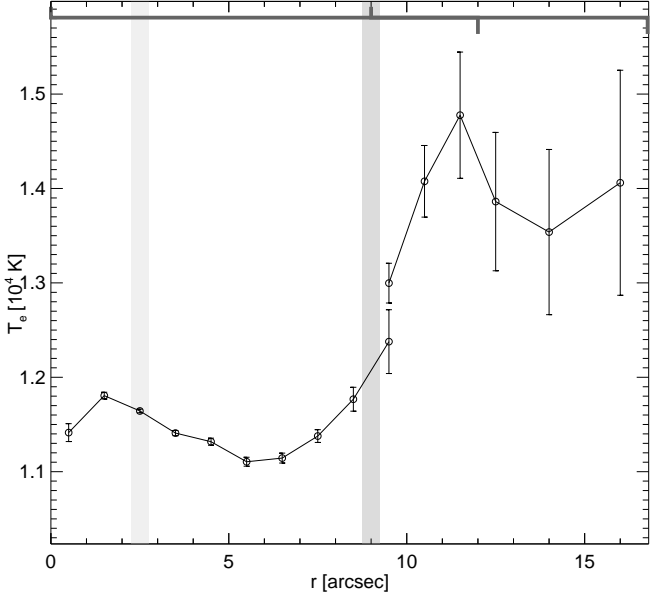}
\caption{Radial temperature structure of {\rIC}, cf.\@ Sect.~\ref{sec:resIC3568T}. For further details of the figure see the caption of Fig.~\ref{fig:resNGC7662Te}.}
\label{fig:resIC3568Te}
\end{figure}

We also derived a temperature structure from the radially binned data, that is shown in Fig.~\ref{fig:resIC3568Te}, including the inner halo; error bars are notably larger in the halo than in the central nebula. The choice of concentric arcs to bin the spaxels in the region inside the shell is, as the two-dimensional map in Fig.~\ref{fig:resIC3568Temap} shows, not optimal (but better than for {\rNGCblue}). Compared to the temperature structures of {\rNGCblue} and {\rNGCeye} (see Figs.~\ref{fig:resNGC7662Te} and \ref{fig:resNGC6826Te}) there is a weaker correlation between the temperature minimum at $r\!\simeq\!6\arcsec$ and the location of the shell-halo transition ($r_{\text{shell}}\!\simeq\!9\arcsec$). {\rIC} also, however, shows an increasing temperature gradient in the region $6\arcsec\!\le\!r\!\le\!12\arcsec$, where the temperature increases from $\Te\!=\!11\,000\,$K to $\Te\!\simeq\!14\,000\,$K.

Previous derivations of the temperature in the central nebula are presented by \citet[$\Te\!=\!10\,800\!\pm\!500\,$K; with revised values calculated by {\rKa}, $\Te\!=\!10\,410\,$K, and \citealt{McKeKa.:96}, $\Te\!=\!9990\,$K]{Ba:78}, {\rKwHe} ($\Tee{\ion{O}{iii}}\!=\!10\,500\!\pm\!5000\,$K, $\Tee{\ion{N}{ii}}\!=\!6900\!\pm\!5000\,$K), \citet[$\Tee{\ion{O}{iii}}\!=\!11\,370\!\pm\!200\,$K and $\Tee{Ba}\!=\!9300\!\pm\!900\,$K]{ZhLiWe.:04}. Most of these values are lower than what we calculated for the central nebula. The temperature map of \citeauthor{ReWo:82} meanwhile shows a larger value in the center (by about $1000\,$K).

\subsubsection{Electron density}\label{sec:resIC3568ne}
A review of electron density measurements in the literature shows large differences: \citet[$\nee{\ion{O}{ii}}\!=\!10\,000\pm9000\,\cbm$]{Ba:78}, {\rSK} ($\nee{\ion{O}{ii}}\!=\!3390\,\cbm$ and $\nee{\ion{S}{ii}}\!=\!3090\,\cbm$, that were derived using observations of previous authors), {\rKwHe} (measure $\nee{\ion{S}{ii}}\!=\!900\!\pm\!700\,\cbm$ $4\arcsec\,$ north of the CS), \citet[$\nee{\ion{O}{ii}}\!=\!2140^{+260}_{-230}\,\cbm$, $\nee{\ion{S}{ii}}\!=\!1550^{+850}_{-620}\,\cbm$, $\nee{\ion{Ar}{iv}}\!=\!1510^{+230}_{-200}\,\cbm$, and $\nee{\ion{Cl}{iii}}\!=\!39^{+220}_{-0}\,\cbm$]{WaLiZh.:04}, and \citet[$\log\nee{Bal}\!=\!3.8\!\pm\!0.2$]{ZhLiWe.:04}. Our observations of [\ion{Ar}{iv}]$\,\lambda4711$ are too inaccurate to make a density estimate -- the actual density also appears to be below the conservative density sensitive range of [\ion{Ar}{iv}] ({\rSK}; $\log\nee{\text{\ion{Ar}{iv}}}\!\la\!3.3$).

\begin{figure}
\includegraphics{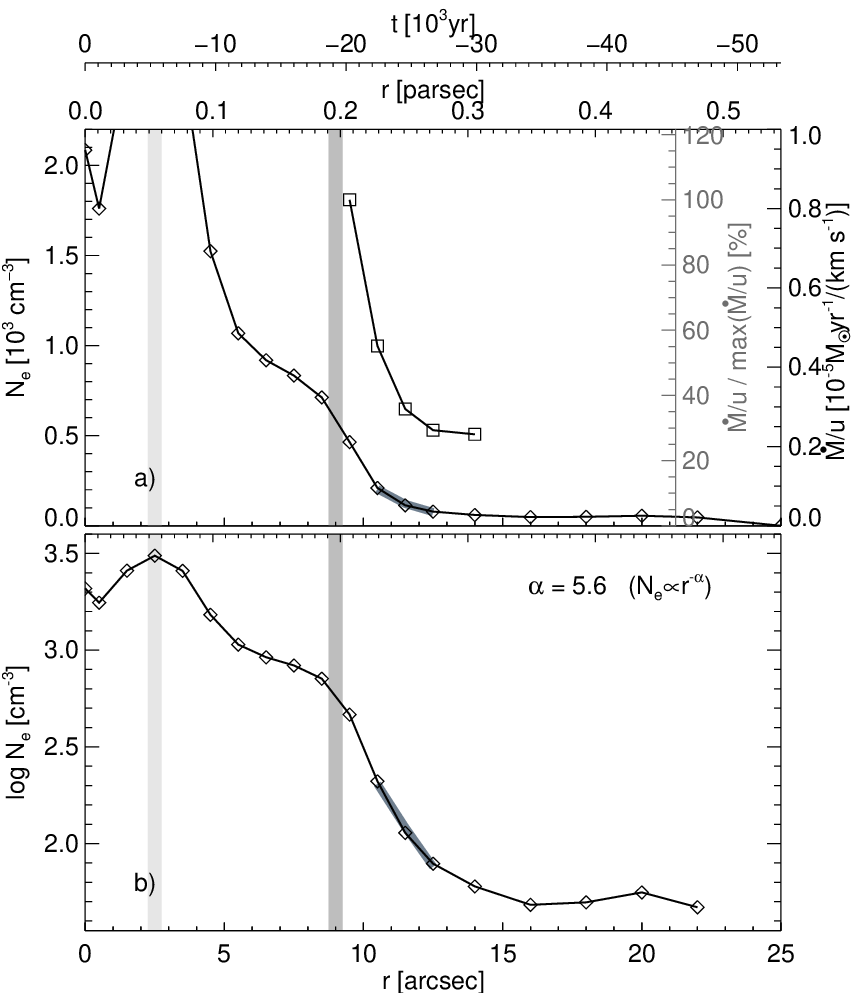}
\caption{Radial density structure of {\rIC}, cf.\@ Sect.~\ref{sec:resIC3568ne}. The thick gray line shows a fit of the density to the power law $\ne\!\propto\!r^{-\alpha}$ for the region $10\farcs5\!\le\!r\!\le\!12\farcs5$. For further details of figure properties see Fig.~\ref{fig:resNGC7662ne}.}
\label{fig:resIC3568ne}
\end{figure}

Instead we calculated a radial density structure using H$\beta$, with the assumptions of $\ne(\text{core})\!\simeq\!2000\,\cbm$ and $d\!=\!4.3\,$kpc, see Fig.~\ref{fig:resIC3568ne}. This density is less accurate for the outer regions ($r\!\ga\!20\arcsec$), since we did not cover the full radial extent of the halo.

\subsection{{\roMtt} -- PN\,G147.8+04.1}\label{sec:resM22}
Existing studies of this small and distant nebula mostly discuss its geometrical shape and distance. For instance, \citet[][and references therein]{CaKaSt:92} measure a diameter of the inner nebula of 7\farcs0. \citet{Zh:95} finds the same size and derives (and quotes) distances of $d\!=\!3.9$--$4.4\,$kpc. Moreover, {\rGVM} \citep{MaGuSt.:96} find the size of the inner nebula to be $5\farcs3\times4\farcs0$, and measure a halo in H$\alpha$ of the proportions $17\farcs4\times14\farcs9$. An [\ion{O}{iii}] image of the nebula is provided by \citet{Ch:89}. \citet{AlKe:87}, finally, calculate an effective temperature $\teff\!=\!80\,000\,$K of the CS.

Values of the extinction coefficients are derived by \citet[$c\!=\!1.25\!\pm\!0.15$]{Ka:83}, \citet[$c\!=\!1.58$]{Ka:85}, \citet[$c\!=\!2.0$]{AlKe:87}, and \citet[$c\!=\!1.46$]{TyAcSt.:92}. Using the same procedure as with {\rNGCblue} we first calculated a map of the extinction coefficient in the central tile. The average of this map is $c\!=\!1.50\pm0.25$, that agrees well with the literature values. The map is used to deredden the data in the blue wavelength range for the central tile, before binning. In comparison we used the average value in tiles 2 and 3, and for all data in the red wavelength range.

The two spaxel binning maps that we used in the blue and red wavelength ranges are illustrated in Fig.~\ref{fig:resspxmapM22}. These maps differ due to the reverse shift of the DAR masks for wavelengths shorter and longer than the reference wavelength $\lambda_{\text{ref}}\!=\!5050\,$\AA. Several emission lines, such as [\ion{N}{ii}]\,$\lambda\lambda6548,\,6583$ and [\ion{S}{ii}]\,$\lambda\lambda6717,\,6730$, are predominantly present in ansae on the south side of the nebula (see the inset in Fig.~\ref{fig:obsimages}c and also fig.~1d of {\rGVM}). In order to not affect the halo measurements when binning the spaxels we masked these regions in tiles 1 and 2.

\subsubsection{Surface brightnesses}\label{sec:resM22sb}
\begin{figure*}
\includegraphics[width=18.3cm]{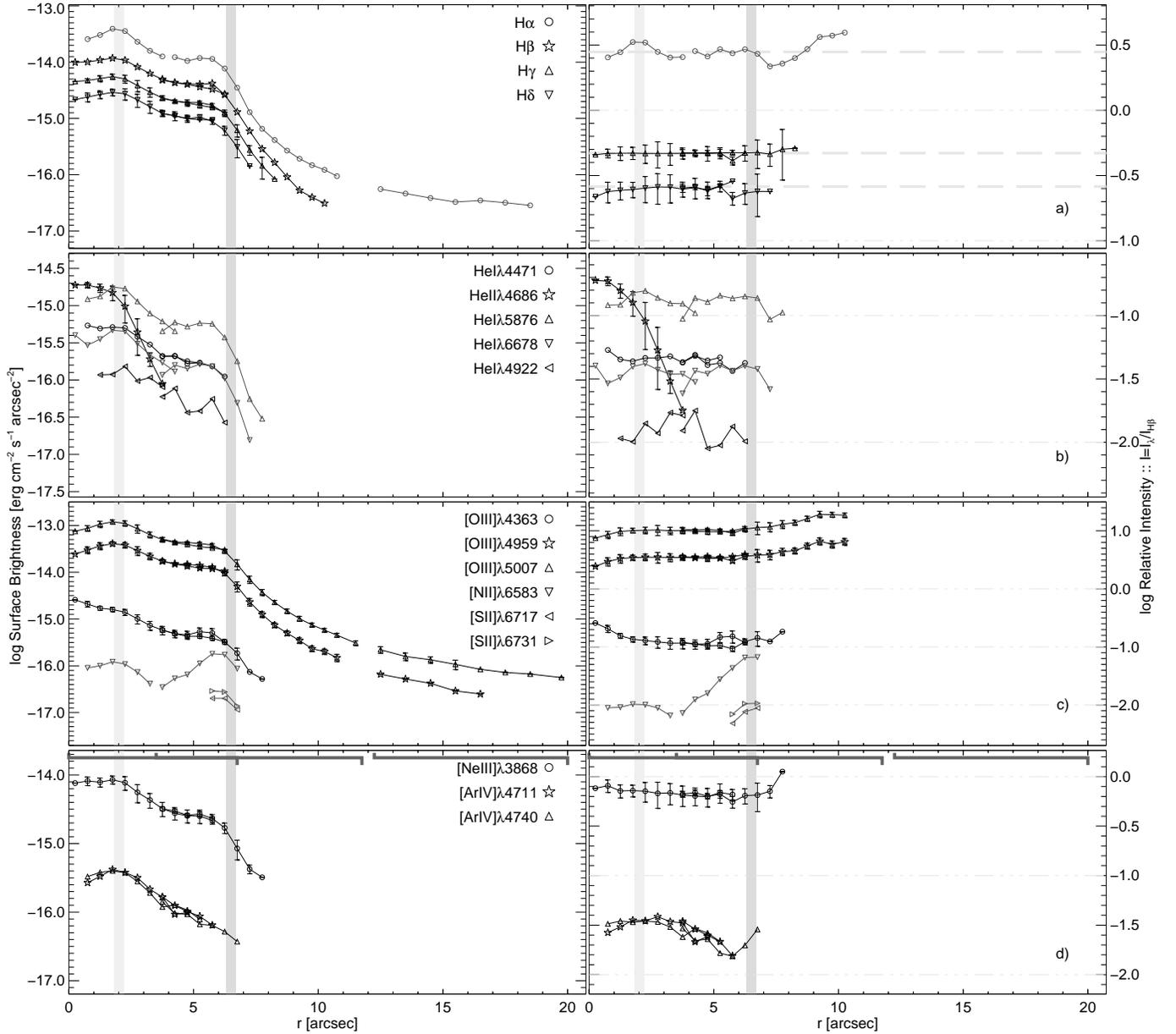}
\caption{Radial surface brightness structures of {\rMtt}, cf.\@ Sect.~\ref{sec:resM22sb}. From the top the panels show 18 emission lines of seven elements: {\bf a)} hydrogen, {\bf b)} helium, {\bf c)} oxygen, nitrogen, and sulfur, and {\bf d)} neon and argon. Data of lines drawn in black in the blue wavelength range were observed 04.10.2007, and data of lines drawn in gray were observed 05.10.2007. Vertical gray bars indicate the location of the rim at $r_{\text{rim}}\!\simeq\!2\farcs0$, and the shell-halo transition at $r_{\text{shell}}\!\simeq\!6\farcs5$. The horizontal dashed gray lines in the right hand side panel {\bf a} indicate line ratios for Case B recombination theory at $\Te\!\approx\!12\,000$\,K. For further details of figure properties see Fig.~\ref{fig:ressbNGC7662}.}
\label{fig:ressbM22}
\end{figure*}

Radial surface brightness structures, that we derived using the binned arcs, are presented in Fig.~\ref{fig:ressbM22}. We determined the locations of the rim and the shell-halo transition from the radial structures of the hydrogen and [\ion{O}{iii}] lines, as $r_{\text{rim}}\!\simeq\!2\farcs0$ and $r_{\text{shell}}\!\simeq\!6\farcs5$. The two emission lines [\ion{O}{iii}]$\,\lambda\lambda4959,\,5007$ were present out to a radius of $r\!=\!17\arcsec$ and $r\!=\!20\arcsec$ from the CS, respectively. In comparison, H$\beta$ was detected out to $r\!=\!8\farcs5$, beyond this radius this line is too weak to be detected. The resulting halo diameter of $D\!=\!40\arcsec$ is consequently $2.3$--$2.7$ times larger than the halo size reported by {\rGVM} (as specified above) and also 2.7 times larger than the diameter \citet{PhRa:06} measure in the $J$ band ($D\!=\!15\arcsec$, see fig.~6 therein). Furthermore, all intensity ratios we measured here between $r\!=\!8\farcs0$ and the bright center are 25--100, which are a factor of ten larger (and more) than the ratios of 2--5 found by {\rGVM}. Finally, from a scattered light test, that we made following the method described in Sect.~\ref{sec:dataanalysisstray}, and using [\ion{O}{iii}]\,$\lambda5007$, we find graphs that are very similar to those shown in Fig.~\ref{fig:dataanalysisstray} for {\rNGCblue} (although with the smaller radii of {\rMtt}). The measured halo of {\rMtt} is thus real.

Showing the, to a good approximation, spherical symmetry of the structure two-dimensional surface brightness maps of four selected emission lines are illustrated in Fig.~\ref{fig:resM22lmap}. The ansae, mentioned in Sect.~\ref{sec:resM22}, are not clearly visible in the presented lines, with the exception of the slightly increased value seen in the S-E corner of the middle tile (right hand side panels in the figure).

\subsubsection{Electron temperature}\label{sec:resM22T}
\begin{figure}
\includegraphics{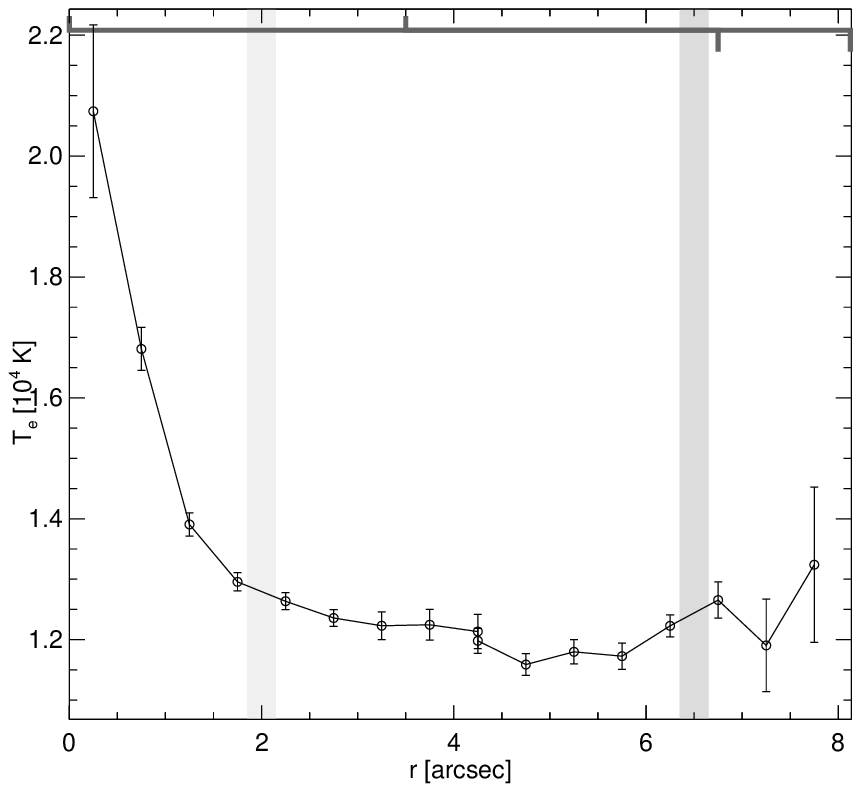}
\caption{Radial temperature structure of {\rMtt}, cf.\@ Sect.~\ref{sec:resM22T}. For further details of figure properties see Fig.~\ref{fig:resNGC7662Te}.}
\label{fig:resTeM22}
\end{figure}

We calculated a temperature using all three diagnostic [\ion{O}{iii}] lines, where the auroral line [\ion{O}{iii}]$\,\lambda4363$ could be measured out to a distance $r\!=\!7\farcs25$, i.e.\@ barely reaching into the halo. The resulting radial structure is presented in Fig.~\ref{fig:resTeM22}. Note that the temperature is very high in the center, $\Te\!\simeq\!20\,000\pm900\,$K, and much lower (and more or less constant) in the remaining parts, $\Te\!\simeq\!12\,000\pm200\,$K. This average value is in perfect agreement with the value of \citet[][$\Te\!=\!12\,000\,$K]{AlKe:87}. Using infrared observations \citet{PhRa:06} note that the index $\left(J-H\right)_0$ decreases systematically away from the CS of {\rMtt}. The authors discuss the possibility to explain such a decrease in terms of hot dust continua. There is in the presented data no clear indication of an increasing temperature in the halo.

\subsubsection{Electron density}\label{sec:resM22n}
\begin{figure}
\includegraphics{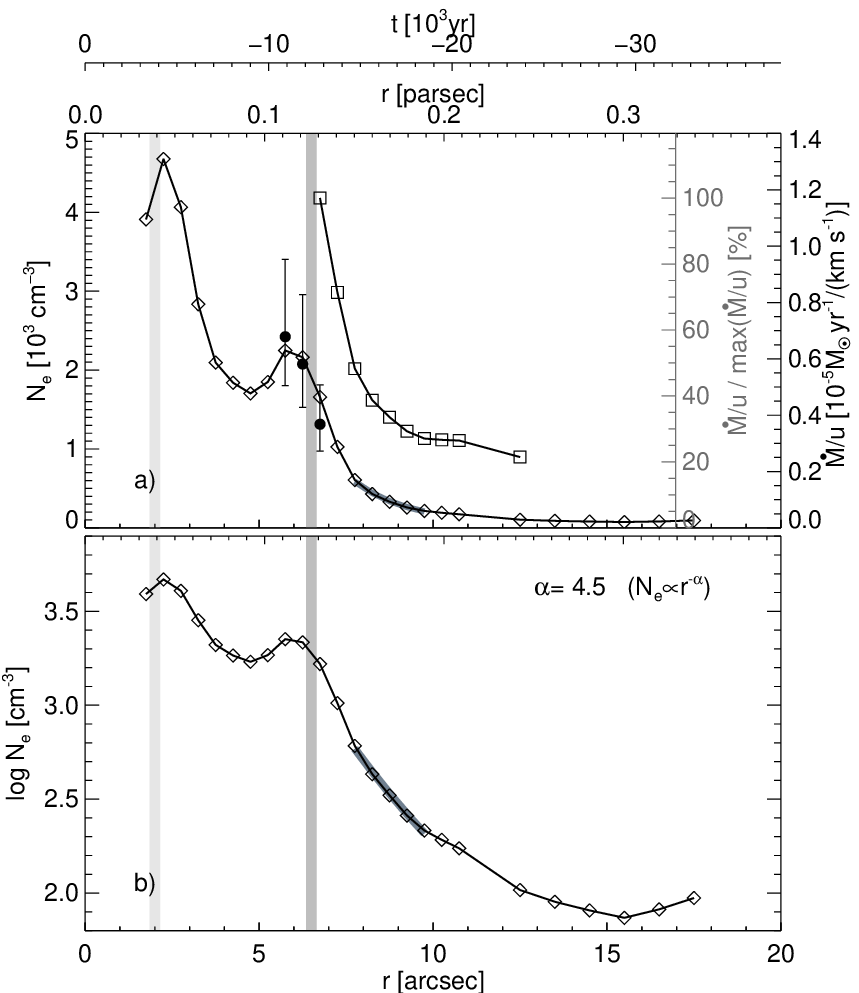}
\caption{Radial density structure of {\rMtt}, cf.\@ Sect.~\ref{sec:resM22n}. The thick gray line shows a fit of the density to the power law $\ne\!\propto\!r^{-\alpha}$, for the region $7\farcs5\!\le\!r\!\le\!9\farcs5$. Bullets with error bars show individual measurements of \nee{\ion{S}{ii}}. For further details of figure properties see Fig.~\ref{fig:resNGC7662ne}.}
\label{fig:resM22ne}
\end{figure}

Due to a comparatively low spectral resolution of the observations we could not accurately deblend the adjacent lines [\ion{Ar}{iv}]\,$\lambda4711$ and \ion{He}{i}\,$\lambda4713$ (where the helium line is found to be the weaker). Instead we measured the sulfur line ratio [\ion{S}{ii}]$\,\lambda\lambda6717/6731$ at $r\!=\!5\farcs75$, 6\farcs25, and 6\farcs75, and found the ratios $0.698\pm0.059$, $0.728\pm0.066$, and $0.829\pm0.073$, respectively. These ratios give the densities $\nee{\ion{S}{ii}}\!=\!2420^{+980}_{-620}$, $2080^{+880}_{-550}$, and $1310^{+500}_{-340}\,$\cbm. In comparison \citet{AlKe:87} measure $\nee{\mbox{\ion{s}{ii}}}\!=\!5000\,\cbm$ in the central nebula.
Additionally we calculated a radial density structure using the Abel transform with H$\alpha$, that we measured throughout most of the halo. When scaling the density structure we assumed a distance of $d\!=\!4\,$kpc. As for {\rNGCblue} and {\rIC} we did not measure the outermost part of the halo, which is why the density is more uncertain for $r\!\ga\!16\arcsec$. Due to asymmetry in the H$\alpha$-structure we did not calculate a density in the central region either. Both the radial structure and the individual measurements are presented in Fig.~\ref{fig:resM22ne}, along with the mass loss rate (cf.\@ Sect.~\ref{sec:discmassloss}).

\subsection{{\roNGCowl} -- PN\,G148.4+57.0}\label{sec:resNGC3587}
{\rNGCowl} has a thin halo outside an extended bright region. The diameter of the PN proper is $D\!>\!200\arcsec$, and surface brightnesses are low. Moreover, since the effective temperature of the CS is so high ($\teff\!=\!93\,900\!\pm\!5\,600\,$K, \citealt{Na:99}; $\teff\!=\!110\,000\,$K, \citealt{McMeKu:97}), and the luminosity so low ($L\!=\!41$--$148\,L_{\sun}$, \citealt{Ka:83b}; $L\!=\!63\,L_{\sun}$ G\'orny priv.\@ comm.), {\rNGCowl} is suggested to have a so-called recombination halo (\citealt{Ty:86}; \citealt{CoScSt.:00}; {\rCSSP}). The halo, that has an asymmetric shape, suggests an interaction with the surrounding interstellar medium \citep[see, for example,][and references therein; the latter reference also provides a list of distance measurements]{SaBiOr.:85,GuChMa.:03}. We observed the shell-halo transition region east of the CS in order to find out what the detectable radial extents of various high- and low-excitation emission lines are, and also to measure a temperature. The spaxel binning map we used with the observations in the red wavelength range is shown in Fig.~\ref{fig:resspxmapNGC3587}. We did not calculate an extinction coefficient, but instead used the constant value $c\!=\!0.0$ \citep[see, e.g.,][]{KaShKw:90,TyAcSt.:92}.

\subsubsection{Surface brightnesses}\label{sec:resNGC3587sb}
We present radial surface brightness structures, which were calculated using the binned arcs, in Fig.~\ref{fig:ressbNGC3587} for 16 observed emission lines. Observational conditions allowed lines in the red wavelength range to be measured in a larger spatial domain, reaching inwards to $r\!=\!65\arcsec$. The agreement between data from the two different years is good, small deviations should be due to sampling in slightly different regions (see Fig.~\ref{fig:obsimages}d). Compared to the other objects of this study the intensities decrease across the observed region by less than a factor of 100. The shell-halo transition at $r\!\simeq\!105\arcsec$ is less sharp than for other objects (also compare with the east [left] side of the two images in fig.~10 of {\rCSSP}). The line intensity structures of H$\alpha$ and [\ion{O}{iii}]$\,\lambda5007$ agree well with fig.~19 in {\rCSSP} -- the rim and shell-halo transition locations were taken from this reference. Furthermore, as is seen in Fig.~\ref{fig:ressbNGC3587}a (right hand side panel) the agreement with recombination theory is fair for H$\alpha$ and H$\gamma$ where $r<105\arcsec$. H$\delta$ is more noisy. Uncertainties are much larger for all lines in the halo.

\subsubsection{Electron temperature}
{\rKa} and \citet{McKeKa.:96} measure an [\ion{O}{iii}] electron temperature of $\Te\!=\!10\,910\,$K and $\Te\!=\!10\,840\,$K, respectively. Both papers apply updated atomic data on the observed line ratio of \citet[who put the slit on top of the CS]{ToPe:77}. \citet{KwHe:01} measure $\Tee{\mbox{\ion{O}{iii}}}\!=\!10\,600\,$K, $\Tee{\mbox{\ion{N}{ii}}}\!=\!9\,400\,$K, $\Tee{\mbox{\ion{O}{ii}}}\!=\!11\,600\,$K, and $\Tee{\mbox{\ion{S}{iii}}}\!=\!12\,400\,$K (all $\pm10\%$). Using the values of [\ion{O}{iii}]$\,\lambda4363$, in the radial range $85\arcsec\!\le\!r\!\le\!95\arcsec$, we found a more or less constant temperature, which average value is $\Tee{\mbox{\ion{O}{iii}}}\!=\!11\,500\pm240\,$K. A comparison with the values of the central region indicates a close to constant temperature across the nebular surface.

\subsubsection{Electron density}
Two of the line doublets we observed are indicators of the electron density, viz.\@ [\ion{O}{ii}] and [\ion{S}{ii}]. However, the density in the shell and halo is low. So low that the measured line ratios of these two indicators are below the conservative density sensitive limits throughout the observed region ({\rSK}, $\log\nee{\text{\ion{S}{ii}}}\!\la\!2.45$ and $\log\nee{\text{\ion{O}{ii}}}\!\la\!2.4$). Neglecting the lower limit the three innermost arcs, at $66\farcs5\!\le\!r\!\le\!68\farcs5$, give an average value $\nee{\text{\ion{S}{ii}}}\!=\!30\pm4\,\cbm$, where the error is the standard deviation of the three values. Unlike for the other objects of this study we did not derive a density structure using the Abel transform. That procedure relies on a complete ionization in the integrated area; because of the low luminosity of the CS that does not appear to be granted.

\citet{SaBiOr.:85} deduce a mean density for the nebula of $\langle\nee{\text{\ion{O}{ii}}},\nee{\text{\ion{S}{ii}}}\rangle\approx100\,\cbm$. {\rSK} (using observational data from \citealt{Os:60} and \citealt{Bo:74}) likewise find $\nee{\text{\ion{O}{ii}}}\!\simeq\!80\,\cbm$ and $\nee{\text{\ion{S}{ii}}}\!=\!170\,\cbm$. \citet{CuPh:00} present a density map using narrow-band imaging of [\ion{S}{ii}]. In that map a distinct dichotomy is found in a relatively flat density gradient between the west and east sides. The average of their map, $\langle\nee{\mbox{\ion{S}{ii}}}\rangle\!=\!590\,\cbm$, appears unrealistic in view of our new -- and previous -- results. \citet{KwHe:01}, finally, find $\nee{\ion{S}{ii}}\!=\!100\!\pm\!10\,\cbm$.

\subsection{{\roNGCeye} -- PN\,G083.5+12.7}\label{sec:resNGC6826}
While the central regions of {\rNGCeye} are often studied in the literature, this is not the case with the halo. Using the {\tPPAK} IFU we observed the full radial extent of the halo with two pointings ($r_{\text{halo}}\!=\!65\arcsec$). Compared with the object image in Fig.~\ref{fig:obsimages}e (i.e., fig.~13 in {\rCSSP}, also see {\rGVM} and \citealt{HaMoTr.:07}) it is seen that the halo of this object deviates only slightly from a spherical shape, justifying our use of concentric arcs to bin spaxels; the two maps that we used are shown in Fig.~\ref{fig:resspxmapNGC6826}. Note that the large diameter of the PPAK fibers prevents us from resolving the central region inside of the rim, which is why we use the radius given by {\rCSSP}, $r_{\text{rim}}\!=\!6\arcsec$. The geometrical properties of the IFU also make it difficult to get perfectly overlapping arcs of spaxels in the two spatially offset tiles. Compare, for example, the spaxel organization of arc \emph{e} in the two tiles. Intensities measured in the same arc, but in different tiles, could consequently, depending on the object properties, differ appreciably. Finally, because of the fiber separation and size, and since the observations were carried out at a low airmass, we did not consider effects of DAR.

To mention a few references, \citet{KiAr:01} measure an extinction coefficient of $c\!=\!0.25$ $4\arcsec$ NE of the CS. \citet{Ba:88} finds $c\!=\!0.16$ (3\arcsec N, 3\arcsec W), and {\rKwHe} measure $c\!=\!0.00$--$0.36$ at three different central positions. In the central part of the nebula, we found a mean value of the extinction coefficient $c\!=\!0.24\pm0.11$. Because of weak H$\gamma$ emission it could not be well determined for the region outside the shell. Six distance estimates of {\rNGCeye} are, moreover, reported by: {\rMiClWao} ($d\!=\!2.3\,$kpc), \citet[$d\!=\!1.9\,$kpc]{MeKuHe:92}, {\rCSSP} ($d\!=\!1.8\,$kpc), \citet[$d\!=\!3.18\,$kpc]{PaHoMe:04}, \citet[$d\!=\!1.2\,$kpc]{Ph:04}, and \citet[$d\!=\!2.6\,$kpc]{KuUrPu:06}. We used $d\!=\!2.5\,$kpc (see Sect.~\ref{sec:discmassloss}). An effective temperature is given by, e.g., \citet[][$\teff\!=\!50\,000\,$K]{KuMePu.:97} and \citet[][$\teff\!=\!46\,000\,$K]{KuUrPu:06}.

\subsubsection{Surface brightnesses}\label{sec:resNGC6826sb}
\begin{figure*}
\includegraphics{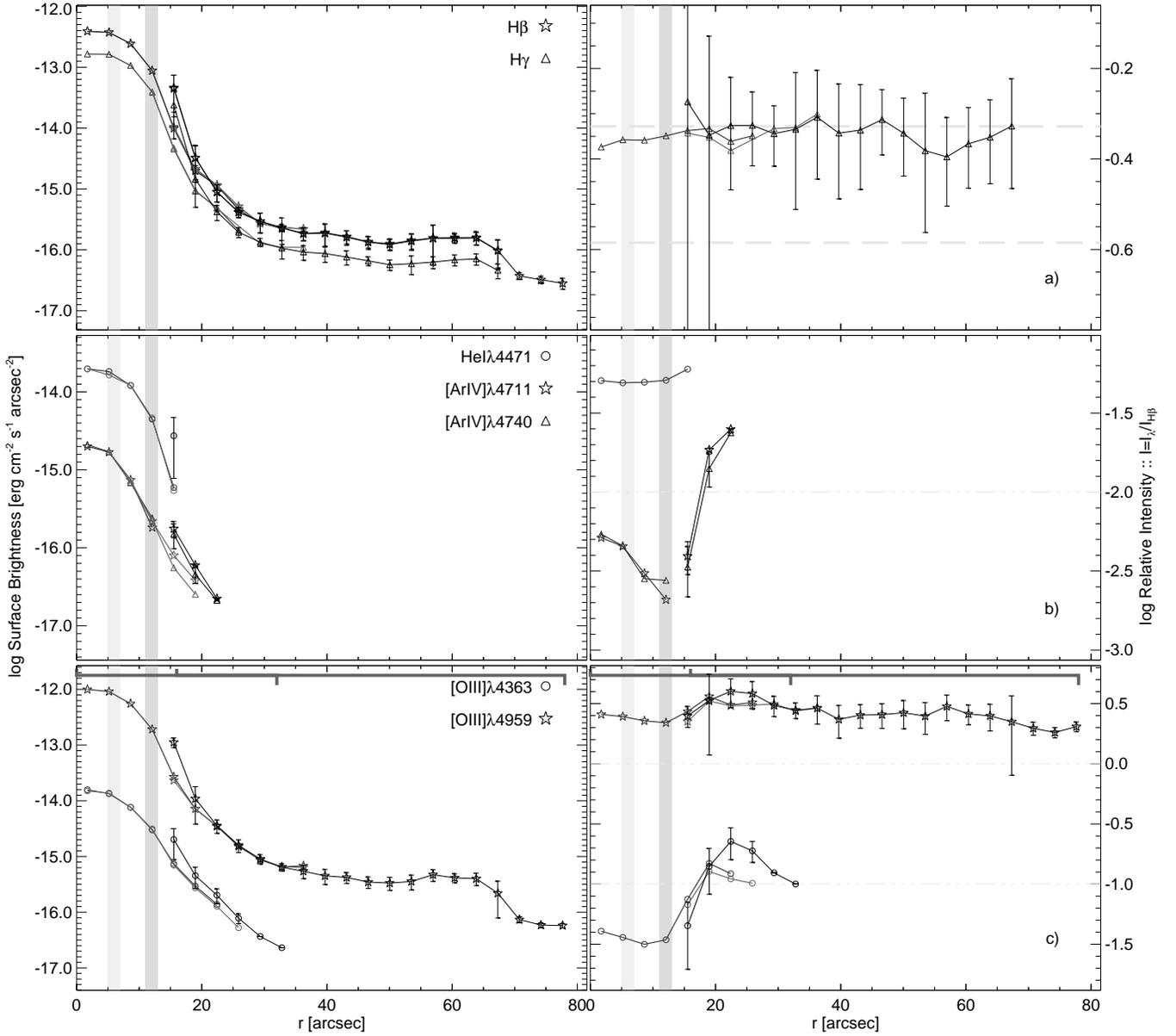}
\caption{Radial surface brightness structures of {\rNGCeye}, cf.\@ Sect.~\ref{sec:resNGC6826sb}. From the top the panels show seven emission line intensities of four elements: {\bf a)} hydrogen, {\bf b)} helium and argon, and {\bf c)} oxygen. For further details of figure properties see Fig.~\ref{fig:ressbNGC7662}. Data of lines drawn in gray were measured in tile 1, and those in black in tile 2 (see Table~\ref{tab:observations}). The vertical gray bars indicate the location of the rim at $r_{\text{rim}}\!=\!6{\arcsec}$, and the shell-halo transition at $r_{\text{shell}}\!=\!12\arcsec$. Note the vertical offset between black and gray lines in the inner halo, for $15\arcsec\!\le\!r\!\le\!25\arcsec$. This offset is due to deviations from symmetry.}
\label{fig:resNGC6826SB}
\end{figure*}

Radial surface brightness structures are presented in Fig.~\ref{fig:resNGC6826SB} for seven emission lines. Note that the full extent of the limb-brightened halo is covered in H$\beta$, H$\gamma$, and [\ion{O}{iii}]$\,\lambda4959$. A comparison of the intensities with the observed radial profiles of {\rCSSP} shows a good quantitative agreement (for H$\alpha$+[\ion{N}{ii}] and $[\ion{O}{iii}]-$\,diffusive light) -- the amount of scattered light in the data is small.

Outside the ``halo boundary'', for radii $r\!\ge\!70\arcsec$, the three outermost measurements of H$\beta$ are underestimated by about 10\,per cent due to a varying continuum level around this emission line. Nevertheless, these data points are shown in order to demonstrate that emission is present. Finally, the helium and argon emission lines \ion{He}{i}$\,\lambda4471$ and [\ion{Ar}{iv}]$\,\lambda\lambda4711,\,4740$ are only detected in the inner part of the halo, out to a radius of about $r\!\simeq\!20\arcsec$.

\subsubsection{Electron temperature}\label{sec:resNGC6826T}
\begin{figure}
\includegraphics{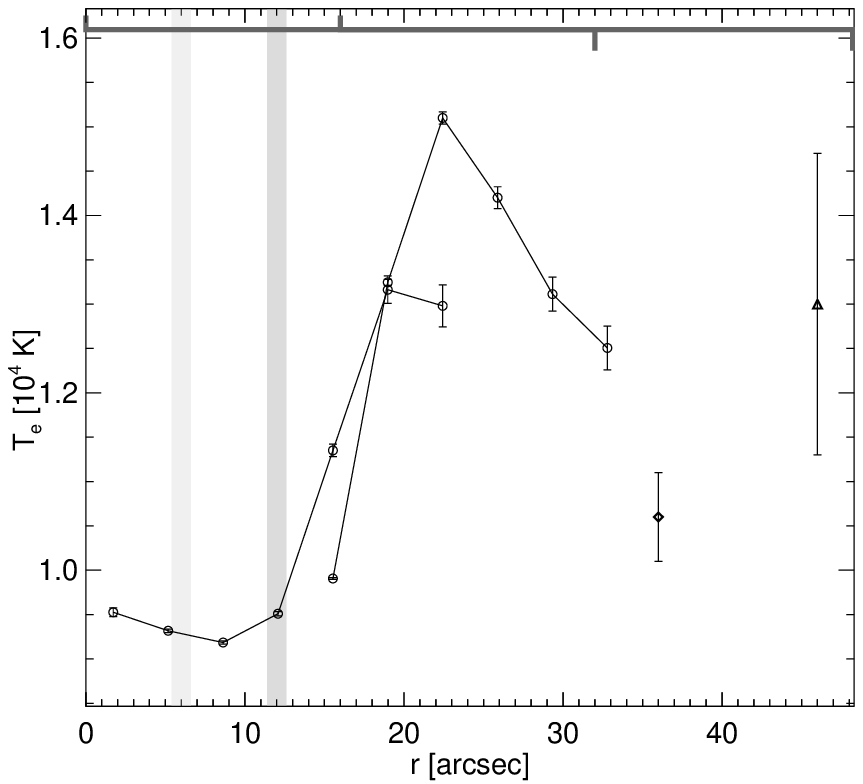}
\caption{Radial temperature structure of {\rNGCeye}, cf.\@ Sect.~\ref{sec:resNGC6826T}. In addition the two halo measurements of {\rMiClWao} ($\triangle$) and {\rMP} ($\diamondsuit$) are indicated. For further details of figure properties see Fig.~\ref{fig:resNGC7662Te}.}
\label{fig:resNGC6826Te}
\end{figure}

We measured an electron temperature out to a distance of $r\!=\!33\arcsec$ from the CS, using the radially binned intensities of [\ion{O}{iii}]$\,\lambda4363$ and [\ion{O}{iii}]$\,\lambda4959$ -- thus covering the inner halo. We present the derived radial temperature structure in Fig.~\ref{fig:resNGC6826Te}.

A steep temperature gradient is seen outwards of the shell, for $r\ga10\arcsec$, where the temperature increases from $\Te\!=\!9200$\,K to $15\,000$\,K within $\Delta r\!\simeq\!12\arcsec$. The maximum is followed by a decreasing temperature in the three outermost arcs, where $\Te\!>\!12\,000$\,K. Two measurements of the halo temperature are reported by \citet[who measure $\Te\!=\!13\,000^{+5000}_{-1700}$\,K at $r\!\approx\!46\arcsec$, hereafter {\rMiClWao}]{MiClWa:89} and \citet[who measure $\Te\!=\!10\,600\pm500$\,K at $r\!\approx\!36\arcsec$, hereafter {\rMP}]{MaPo:89} -- both values are indicated in Fig.~\ref{fig:resNGC6826Te}. These values are measured farther out in the halo than what was possible with the data presented here. The value of {\rMP} is about $2000\,$K lower than we measured at $r\!=\!33\arcsec$.

{\rJQA} present a temperature (and density) map of the central nebula. The agreement with our data is good. In addition to the halo values {\rMiClWao} and {\rMP} measure $\Te=10\,400\pm750$\,K and $\Te=9\,100\pm100$\,K in the core, respectively. Both these papers report on the apparent increasing gradient using the two respective values of the central parts and the halo. Additional values are reported by \citet[who use data of \citealt{Ka:86}, $\Te\!=\!11\,110$\,K]{McKeKa.:96}, {\rKwHe} ($\Te\!=\!8800$--$9000\pm500$\,K), and  \citet[$\Te\!=\!9350\pm200$\,K]{ZhLiWe.:04}.

\subsubsection{Electron density}\label{sec:resNGC6826ne}
To mention a few references {\rJQA} provide a density map of the central nebula, using the sulfur line doublet, $\nee{\ion{S}{ii}}$. Using the same lines \citet{Ba:88} and {\rKwHe} calculate a density at six and three different positions, respectively. {\rMP} determine average values from the H$\beta$ flux, $\ne(\text{core})\!=\!2000\,$\cbm and $\ne(\text{halo})\!=\!20\,\cbm$. \citet{WaLiZh.:04} calculate $\log\nee{\text{\ion{O}{ii}}}\!=\!3.27^{+0.06}_{-0.06}$, $\log\nee{\text{\ion{S}{ii}}}\!=\!3.28^{+0.06}_{-0.06}$, $\log\nee{\text{\ion{Cl}{iii}}}\!=\!3.12^{+0.08}_{-0.10}$, and $\log\nee{\text{\ion{Ar}{iv}}}\!=\!3.33^{+0.15}_{-0.21}\,\cbm$. \citet{ZhLiWe.:04} find a value derived from spectra near the Balmer jump, $\log\nee{Bal}\!=\!3.4\pm0.1\,\cbm$. We observed the two density sensitive argon lines, but in this case the helium line \ion{He}{i}$\,\lambda4713$ is of similar strength as [\ion{Ar}{iv}]$\,\lambda4711$, which is why we could not deblend the two lines accurately. Note that the expected electron density also is below the conservative sensitivity limit of the argon line doublet ({\rSK}, $\log\nee{\mbox{\ion{Ar}{iv}}}\!\la\!3.3$).

\begin{figure}
\includegraphics{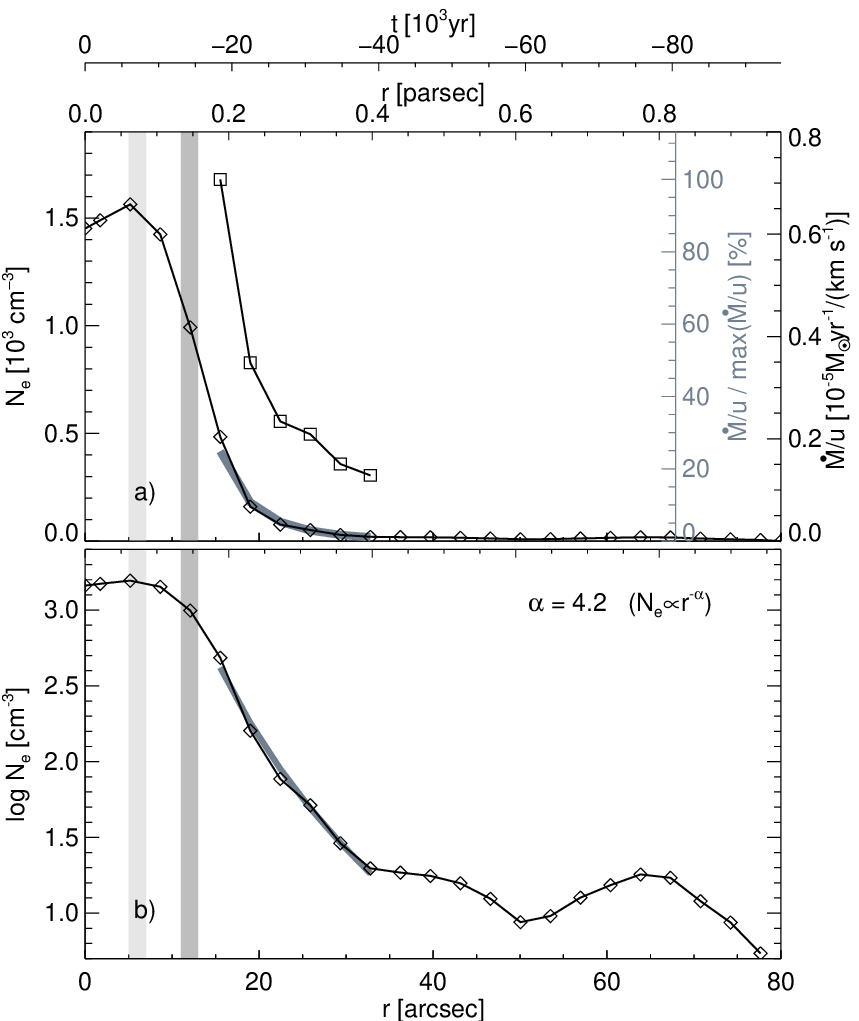}
\caption{Radial density structure of {\rNGCeye}, cf.\@ Sect.~\ref{sec:resNGC6826ne}. The thick gray line shows a fit of the density to the power law $\ne\!\propto\!r^{-\alpha}$ for the region $15\arcsec\!\le\!r\!\le\!33\arcsec$. For further details of figure properties see Fig.~\ref{fig:resNGC7662ne}.}
\label{fig:resNGC6826ne}
\end{figure}

Following the approach of {\rPlSo} for {\rNGCeye} we also used an Abel transform on the radial structure of H$\beta$ in order to calculate a density structure. {\rPlSo} fix the density to $\ne\!=\!1000\,\cbm$ at the radius $r\!=\!12\arcsec$. We derived a similar structure under the additional assumptions of a distance of $d\!=\!2.5\,$kpc and a temperature of $\Te\!=\!12\,000$\,K. With these values we also get $\ne(r\!=\!12\arcsec)\!=\!1000\,\cbm$, which at this radius is well within the azimuthal density range reported by {\rJQA} ($800\!\la\!\nee{\text{\ion{S}{ii}}}\!\la\!2000\,$\cbm). The result is shown in Fig.~\ref{fig:resNGC6826ne}, also see Sect.~\ref{sec:discmassloss} for a discussion on the derived mass loss rate. 

\section{Discussion}\label{sec:discussion}
In this section we first discuss the appearance of hot halos in Sect.~\ref{sec:dischalos}. Thereafter we derive mass loss rates and core and halo masses for three of the objects of this study in Sect.~\ref{sec:discmassloss}.

\subsection{Occurrence and understanding of hot halos}\label{sec:dischalos}
A knowledge of the electron temperature structure is important when determining the ionization structure and abundances of elements in PNe, including the halo. In our halo observation data we detected emission lines of hydrogen, oxygen, neon, and occasionally also helium and argon in the inner halo. Halo temperatures can unlikely be determined by any other method than through the oxygen [\ion{O}{iii}] emission lines.

{\rMiClWao} ({\rNGCeye}), {\rMP} \citep[{\rNGCeye} and NGC\,6543, also see][for the latter]{MaViRi.:07}, {\rMiClWa} ({\rNGCblue}), and \citet[NGC\,3242]{MoRoSc.:05} present temperatures calculated at one position in the core and one in the halo for the respective object. Radial temperature structures which seem to indicate high halo temperatures for {NGC\,1535} and {NGC\,3242} are presented by \citet[see figs.\@ 5 and 13 therein]{KrCo:05}. Compared to these studies we derived densely sampled temperature structures for four out of five nebulae, reaching from the PN core and out into the halo. In three of the four temperature structures we found a hot halo: {\rNGCblue} (Fig.~\ref{fig:resNGC7662Te}), {\rNGCeye} (Fig.~\ref{fig:resNGC6826Te}), and {\rIC} (Fig.~\ref{fig:resIC3568Te}). The figures show that the region is narrow, where the temperature increases from the cooler PN core to the hotter halo. Note that de-projected (real) temperature structures are expected to be different, and temperatures can be even higher. Very high temperatures could be important, as H$\alpha$ becomes collisionally excited when temperatures are higher than $20\,000\,$K \citep{St:02}. 

Early studies, which are based on hydrostatic model formulations, explain hot halos as a consequence of radiation hardening or shock heating mechanisms (see, for example, {\rMP}; {\rMiClWao}; {\rMiClWa}). Due to high densities a PN core is in general in thermal equilibrium, but the tenuous halo is not. Non-equilibrium models, which account for hydrodynamic effects, are first suggested to explain high temperatures in the halo by \citet[also see \citealt{Ma:95} and \citealt{MeFr:95}; the models are based on the formalism introduced by \citealt{Ma:94} and \citealt{MaSz:97}]{Ma:93} and \citet{Ty:03}. With such models the ionization front, which is responsible for the high temperature in the halo, moves through the envelope in a short time period (about $10^2$--$10^3\,$ years long). These models suggest that high temperatures in the halo are a fairly common, although transient, phenomenon.

\begin{figure}
\centering
\includegraphics{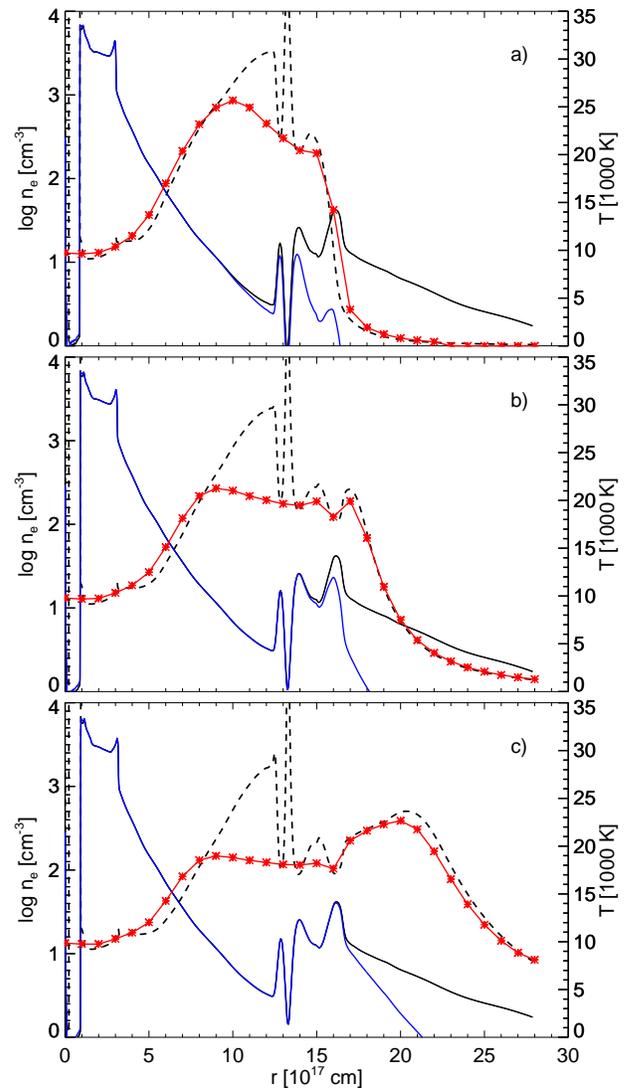}
\caption{A radiation hydrodynamic model sequence showing the evolution of a radial PN structure as the halo is being ionized. Between panels a and c the model ages by 125 years, and the effective temperature and the luminosity of the CS change from $\teffe{a}\!=\!76\,800\,$K and $L_{\text{a}}\!=\!5850\,\text{L}_{\sun}$ to $\teffe{c}\!=\!80\,200\,$K and $L_{\text{c}}\!=\!5790\,\text{L}_{\sun}$. The four lines show: the total number density of atomic nuclei (black solid line), the electron density (blue solid line), the real electron temperature (dashed line), and the spectroscopic temperature (red line connected with asterisks). The density ordinates are logarithmic (left hand side axes). The halo is the region between the shell-halo transition, at $r_{\text{shell}}\!\simeq\!3\times10^{17}\,$cm and $r_{\text{halo}}\!\simeq\!12\times10^{17}\,$cm. The radial domain $12\times10^{17}\!\le\!r\!\le\!18\times10^{17}\,$cm is the wind interaction region of the last thermal pulse. Mass loss occurring before the last thermal pulse is found where $r\!\ge\!18\times10^{17}\,$cm.}
\label{fig:discmodels}
\end{figure}

\begin{figure}
\centering
\includegraphics[width=8.8cm]{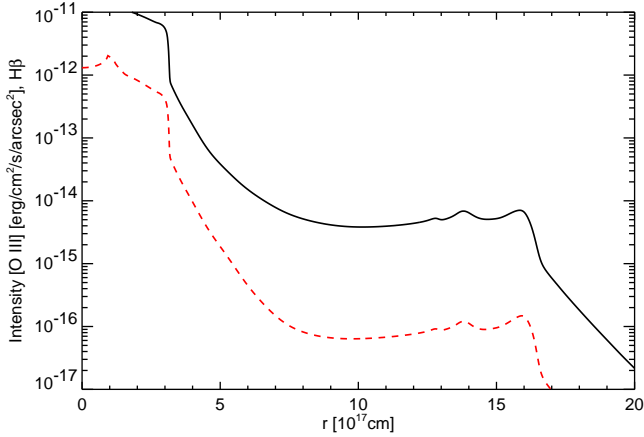}
\caption{The surface brightness of H$\beta$ (red dashed line) and [\ion{O}{iii}]$\,\lambda5007$ (solid line) calculated using the physical structure shown in Fig.~\ref{fig:discmodels}c. For further details see Sect.~\ref{sec:dischalos}.}
\label{fig:discmodsb}
\end{figure}

In Figs.~\ref{fig:discmodels}a--c we show a sequence of radial PN structures that were calculated using a time-dependent radiation hydrodynamic model, where $\teff\!\simeq\!80\,000\,$K and $L\!\simeq\!6000\,\text{L}_{\sun}$. The starting model, for the calculations of the hydrodynamic evolution through the PN stage, is taken from \citet[see fig.~21 therein]{StSzSc:98}. We selected this model since it shows how the R-type ionization front moves through the former AGB wind, thereby giving rise to an observable halo. We also wanted to show qualitatively how the high temperature occurs during the ionization of the halo -- the model is not a fit to the observation data. An R-type ionization front is moving outwards and ionizes the halo, see the blue solid line that shows the electron density. In only 125 years the spectroscopic\footnote{The spectroscopic temperature is derived from the computed line profiles of [\ion{O}{iii}] as a function of the distance from the CS.} electron temperature in the central halo (where $7\!\la\!r\!\la\!12\times10^{17}\,$cm) changes from $\Te\!\simeq\!26\,000\,$K, in Fig.~\ref{fig:discmodels}a, to $\Te\!\simeq\!19\,000\,$K, in Fig.~\ref{fig:discmodels}c, see the red line marked with asterisks. Note that in comparison to the spectroscopic temperature, which is projected onto the line of sight, the real temperature (dashed line) can be about $10^4\,$K larger in the same region. The spectroscopic temperature in the halo is strongly affected by the cool and dense wind interaction region at $12\!\la\!r\!\la\!18\times10^{17}\,$cm, causing the rapid decrease. A use of the spectroscopic temperature to calculate abundances would therefore give incorrect values. Figure~\ref{fig:discmodsb} shows the surface brightnesses of H$\beta$ and [\ion{O}{iii}]$\,\lambda5007$. Note the intensity maximum in the halo for $13\!\la\!r\!\la\!17\!\times10^{17}\,$cm, that is also visible in Fig.~\ref{fig:resNGC6826SB}d for {\rNGCeye}.

\subsection{Deriving mass loss rates for winds at the tip of the AGB}\label{sec:discmassloss}
Starting with an electron density (\ne), that we derived using an inverse Abel transform, we first calculate a mass density ($\rho$). With a mass density structure and a velocity structure it is then straightforward to calculate a radially dependent mass loss rate ($\dot{M}$). Under the three assumptions of: full ionization, a simple chemistry of 90\%\,H and 10\% (singly ionized) He, and a velocity structure $u(r)$ we get,
\begin{eqnarray*}
\rho\left(r\right)\!=\!\big[0.9\!\times\!1+0.1\!\times\!4\big]m_{\text{u}}\ne\!\left(r\right)\;\;\mbox{and}\;\;\dot{M}\left(r\right)=4\pi r^2\rho\left(r\right)u\left(r\right),
\end{eqnarray*}
where $m_{\text{u}}$ is the atomic mass constant. The velocity structure is in general not known, which is why we instead provide the following property,
\begin{eqnarray*}
\dot{M}\left(r\right)/u\left(r\right)=4\pi r^2\rho\left(r\right),
\end{eqnarray*}
that can be multiplied by any radially dependent velocity to get the mass loss rate. In this study we assumed in every case a constant velocity, $u\left(r\right)\!=\!10\,\kms$.
We present radial mass loss rate structures -- that in every figure are indicated by the line that is connected by square symbols -- for {\rNGCblue} (Fig.~\ref{fig:resNGC7662ne}), {\rIC} (Fig.~\ref{fig:resIC3568ne}), {\rMtt} (Fig.~\ref{fig:resM22ne}), and {\rNGCeye} (Fig.~\ref{fig:resNGC6826ne}). We did not consider {\rNGCowl} since it is an old object that is not expected to contain any information of the AGB wind anymore. 
Note that densities and masses depend on the assumed distance ($d$) to the object: mass loss rates scale with the distance as $\dot{M}\!\propto\!d^{3/2}$ and integrated masses as $M\!\propto\!d^{5/2}$.

\begin{table*}
\caption{Calculated mass loss rates and core and halo masses, cf.\@ Sect.~\ref{sec:discmassloss}. Object names and distances ($d$) are given in the first two columns. Columns 3--8 specify six properties: the effective temperature (\teff), the radial range $\Delta r\!=\!r_{\text{min}}$--$r_{\text{max}}$ where we calculated a mass loss rate, the range of exponents $\alpha$ (of fits to $\ne\!\propto\!r^{-\alpha}$) in the radial range $\Delta r$ where $\alpha_{\text{min}}$ was calculated just inside $r_{\text{max}}$ and $\alpha_{\text{max}}$ just outside $r_{\text{min}}$, the maximum mass loss rate in the same range ($\dot{M}_{\text{max}}$), the maximum-to-minimum mass loss rate ratio in the same range ($\xi$), and the resulting mass of the core and the halo in the region where we calculated a density ($M_{\text{core}}$, $M_{\text{halo}}$). The halo mass is printed in italics where the full radial extent of the halo was not covered. The subscript index denotes the age of the radial extent where we integrated a halo mass (specified in thousands of years). A constant outflow velocity of $u\!=\!10\,\kms$ was assumed throughout all objects. The comma separated core and halo masses in Col.~9 ($M_{\text{core},\text{ref}}$, $M_{\text{halo},\text{ref}}$) are based on the values given in the respective reference, that are specified in Col.~11. The original masses, that used the distances specified in Col.~10 ($d_{\text{ref}}$), were scaled to use our distances ($d$, Col.~2). References: (1) \citet{HiBaGr:85}, (2) {\rMiClWao}, (3) {\rMP}, (4) {\rPlSo}, and (5) {\rMiClWa}.}\label{tab:discmassloss}
\begin{tabular}{rcrcccccccc}
\noalign{\smallskip}
\hline\hline\\[-2ex]
\multicolumn{1}{c}{Object} & $d$ & \multicolumn{1}{c}{\teff} & $r_{\text{min}}$--$r_{\text{max}}$ & $\alpha_{\text{max}}$--$\alpha_{\text{min}}$ & $10^5\dot{M}_{\text{max}}$ & $\xi$ & $M_{\text{core}}:M_{\text{halo}}$ & $M_{\text{core},\text{ref}}:M_{\text{halo},\text{ref}}$ & $d_{\text{ref}}$ & Ref.\\
       & $\left[\text{kpc}\right]$ & \multicolumn{1}{c}{$\left[10^3\,\text{K}\right]$} & [arcsec]& & $\left[M_{\sun}\text{yr}^{-1}\right]$ & & $\left[M_{\sun}\right]$ & $\left[M_{\sun}\right]$ & $\left[\text{kpc}\right]$ \\
\multicolumn{1}{c}{(1)} & (2) & \multicolumn{1}{c}{(3)} & (4) & (5) & (6) & (7) & (8) & (9) & (10) & (11)\smallskip\\\hline\\[-1ex]
{\rNGCblue} & 1.7 & 100 & 15--36 & 6.6--2.4 & 10  & 6.7 & $<\!0.64:\mathit{0.53}_{17}$ & 0.053:0.32, 0.19:0.26 & 1.15, 1.5 & (1), (5)\\
{\rIC}      & 4.3 &  55 & 16--33 & 7.9--2.4 & 7.4 & 3.6 & $0.77:\mathit{0.91}_{33}$ \\
{\rMtt}     & 4.0 &  80 & 10--14 & 7.3--2.6 & 12  & 4.7 & $0.50:\mathit{0.88}_{25}$ \\
{\rNGCeye}  & 2.5 &  46 & 6.8--13 & 5.5--3.4 & 7.1 & 5.5 & $0.50:1.9_{76}$ & 0.56:1.8, 0.52:1.2, & 1.02, 2.3, & (1), (2),\\
            &     &     & &          &     &     &                 & 0.59:1.1, 0.50:1.3 & 1.0, 1.54 & (3), (4)\\\hline
\end{tabular}
\end{table*}

Only parts of a halo contain information of the mass loss rate on the AGB. In the region that is close to the shell photoionization processes have affected the wind structure, and in the outer parts of the halo the wind structure has been affected by interactions between the wind and previous mass loss episodes. For three out of the four objects, where we measured density structures, we observed only the inner halo. The resulting ``truncated'' intensity structure of these objects introduces an error into the derived density, that is above a few per cent only in the outermost parts. In the central nebula (i.e., the PN core) an asymmetric structure ({\rNGCblue}) or a low spatial resolution (\rNGCeye) makes the density inaccurate there, which is why our core masses likely are uncertain for these objects.

We present the results of our study in Table~\ref{tab:discmassloss}. In every case, but {\rMtt} (where there is no comparison), we based the distances on the Shklovsky distances (and luminosities) given by G\'orny (priv.\@ comm., see {\rCSSP}). The four PNe are in a similar evolutionary state and should have similar luminosities, and we therefore scaled the distances to get the same luminosity, $L\!\simeq\!6000\,\text{L}_{\sun}$, that corresponds to a CSPN mass of $\simeq\!0.6\,\text{M}_{\sun}$. When we calculated halo masses we used the entire radial extent of the density structure we had outside the shell. We only observed the full radial extent of the halo for {\rNGCeye}, the halo masses of the remaining objects are lower limits only. When we calculated core masses we used the region between the rim and the shell. Our distances are larger than those given in the literature, and our masses are consequently also larger. In Table~\ref{tab:discmassloss} we also compare our core and halo masses with literature values ($M_{\text{core},\text{ref}}$ and $M_{\text{halo},\text{ref}}$). Since adopted distances differ between references we scaled the respective value to use our distances.

Our masses should be in best agreement with {\rPlSo}, since we used the same method as they do. In comparison \citet{HiBaGr:85} assume the density dependence $n\!\propto\!r^{-2}$, and the other three papers use a constant density in the halo. If we truncate the density structure at $r\!=\!70\arcsec$, that is the outer boundary used in {\rPlSo}, we get the halo mass $M_{\text{halo}}\!=\!1.5\,\text{M}_{\sun}$, a value that is in good agreement with $M_{\text{halo},\text{\rPlSo}}\!=\!1.3\,\text{M}_{\sun}$. Compared to {\rPlSo} we, furthermore, did not use the entire halo surface to determine the density structure, and the spatial resolution of our data is lower, in particular in the PN core. We attribute the remaining deviation in the outcome to these differences.

A power law fit of the density with the radius ($\rho\!\propto\!r^{-\alpha}$) should, if the mass loss rate and outflow velocity are constant, give the exponent $\alpha\!=\!2$. For each object we measured several values of $\alpha$ in the same region where we calculated a mass loss rate. The upper and lower limits of these values are presented in Col.~5 of Table~\ref{tab:discmassloss}. In every case $\alpha$ is greater than 2, and increases to $\alpha\!\simeq\!5$--8 in the inner part of the halo -- indicating a marked increase of mass loss during the final outflow stage on the AGB. Fits of $\alpha$ for separate regions are shown in the density plots of the respective object; the value of {\rNGCeye}, $\alpha\!=\!4.2$, agrees well with that derived by {\rPlSo}, $\alpha\!=\!4$. Note, however, that the ionized AGB wind is accelerated by its own pressure gradient and is no longer exactly described by the original power law index $\alpha'$ of the neutral wind \citep[see][]{ScJaSt.:05}.

The property $\xi$, that is the ratio between the minimum and maximum measured mass loss rates, varies between 3.6--6.7. Note that both $\alpha$ and $\xi$ are independent of the distance. Despite the relatively low luminosity (of only $\sim\!6000\,\text{L}_{\sun}$) and increasing effective temperature of the PPNe moving off the tip of the AGB, the maximum mass loss rates average at $\dot{M}_{\text{max}}\!\simeq\!10^{-4}\,\text{M}_{\sun}$. Because of the presence of scattered light, in that part of the halo which lies the closest to the shell, mass loss rates are there over-estimated by up to about 30 per cent (Sect.~\ref{sec:dataanalysisstray}); hence real slopes are less steep. Moreover, a use of shorter distances, which are smaller by a factor of 2, result in a factor 3 lower average mass loss rate. The four mass loss rates are also similar, within a factor of 2. These mass loss rates are important constraints on the theory of wind formation, which so far does not treat winds on the tip of the AGB satisfactorily. Results of earlier studies which indicate an increasing mass loss rate are reported by {\rMP}, {\rPlSo}, and \citet{ScJaSt.:05}, also see \citet{KwVoHr:02}.

\section{Conclusions}\label{sec:conclusions}
Surface brightnesses of emission lines in the halo of a PN can be several thousand times weaker than in the PN core. Therefore it is a demanding task to derive physical properties of the halo. Previous studies of PNe, which focus on the halo, rarely derive physical properties at more than one position. For this paper we wanted to make an observational study of PNe in the galactic disk, where we put the emphasis on the halo and the shell-halo transition regions. Our goal was to derive densely sampled temperature structures extending from the PN core and as far out in the halo as possible. And also to study the final mass loss history on the AGB.

We carried out the observations using the IFS instrument PMAS. Through this approach we could account for several issues -- in our data analysis -- which treatments are critical to the outcome when evaluating steep intensity gradients, such as found in PNe. Next we summarize our results of four such issues. First, in every case, but one (where it was not necessary), we corrected individual spectra in the reduced data for DAR. Depending on observing conditions we thereby avoided errors in derived temperatures of about a thousand Kelvin. Second, in order to save valuable observing time we developed a new method to subtract telluric emission lines from the spectra, instead of using separate sky exposures. A telluric emission line of key importance to our studies is the mercury line Hg$\,\lambda4358$ that often is strong and lies adjacent to the temperature sensitive, and often very weak, oxygen line [\ion{O}{iii}]$\,\lambda4363$. At the rather low spectral resolution of our observations a careless subtraction of Hg\,$\lambda4358$ may lead to temperatures which are several thousand Kelvin too large. Our novel method, where telluric lines and line wings, are fitted with a set of lines of fixed proportions -- together with the object lines -- is useful and general for as long as the telluric and the object lines are resolved.

Third, in order to increase the signal-to-noise of individual emission lines we binned spectra of individual spatial elements on the IFUs along concentric arcs -- under the assumption that the halo structure is spherically symmetric. The arcs were centered on the CS. With this method we extracted intensities at increasing radii -- at a spatial separation and width of $\Delta r\!\ge\!0\farcs5$--$3\farcs6$, depending on the choice of the IFU and line strengths. Fourth, and finally, we made a study of the contribution of scattered light to the data when using PMAS and the 3.5m telescope at Calar Alto. The results showed that some scattered light is present in the inner halo. Its contribution at any radial point is moderate, $\la\!30\%$, and mainly comes from bright nearby regions; from within a distance of about $4\arcsec$ when using the 1\farcs0 IFU. Our observations of faint halos consequently show real structures.

For the extended and old nebula {\rNGCowl} we could determine one value of the electron temperature in the outer parts of the nebula that agrees with earlier values in the literature. Of the remaining four objects in our study, we found that {\rMtt} has a halo radius that is more than twice as large as reported earlier ($r_{\text{halo}}\!\simeq\!20\arcsec$). For {\rIC} we found a halo ($r_{\text{halo}}\!\ge\!25\arcsec$) where this had not been detected earlier. We, moreover, calculated radial temperature structures covering the PN core and parts of the halo for {\rNGCblue}, {\rIC}, {\rMtt}, and {\rNGCeye}. While the structure of {\rMtt} showed a large temperature in the center, and an otherwise constant value, the remaining three structures showed a steep radial temperature gradient with significantly higher values in the halo. Hot halos are described with the help of hydrodynamic models as transients, which in a short time move through the halo. Since matter in halos is rarefied it takes a long time before a equilibrium conditions are established.

We derived electron densities in the shell for {\rNGCblue} (\nee{\ion{Ar}{iv}}) and {\rMtt} (\nee{\ion{S}{ii}}). Using radial surface brightness structures, and assuming a constant radial velocity, we additionally derived density and mass loss rate structures, and masses, for {\rNGCeye}, {\rNGCblue}, {\rIC}, and {\rMtt}. All mass loss rate structures show an increasing outflow in the final stage of mass loss on the AGB. Comparing the four ratios of maximum to minimum mass loss rates these are about 4--7, and within a factor of two the maximum mass loss rate is $\dot{M}_{\text{max}}\!\simeq\!10^{-4}\,\text{M}_{\sun}\text{yr}^{-1}$. A remarkable value in view of the relatively low luminosity of these objects that is close to $6000\,\text{L}_{\sun}$. It would be of great interest to see what the outcome would be of theoretical models of mass loss, which assume the physical properties at the tip of the AGB.

Finally, a logical follow-up of this study is to calculate abundances of the AGB wind in the inner halo. This is, however, only possible for a few elements -- such as helium, oxygen, neon, and maybe also argon -- since intensities are so weak. The results of our study should, moreover, be valuable to comparisons with objects in the halo and the Magellanic Clouds. In a new study we are working on the detection and measurement of halos in a sample of PNe in the Small and Large MCs, in order to study closer the influence of the metalicity on the physical structure. The spatial resolution, which can be achieved with these objects is naturally lower than with galactic disk objects, since the two MCs are more distant.

\begin{acknowledgements}
C.S.\ is supported by DFG under grant number SCHO 394/26.
\end{acknowledgements}

\bibliographystyle{aa}
\bibliography{XPN_Refs,CS_Refs}

\onlfig{24}{
\begin{figure*}
\sidecaption
\includegraphics[width=12cm]{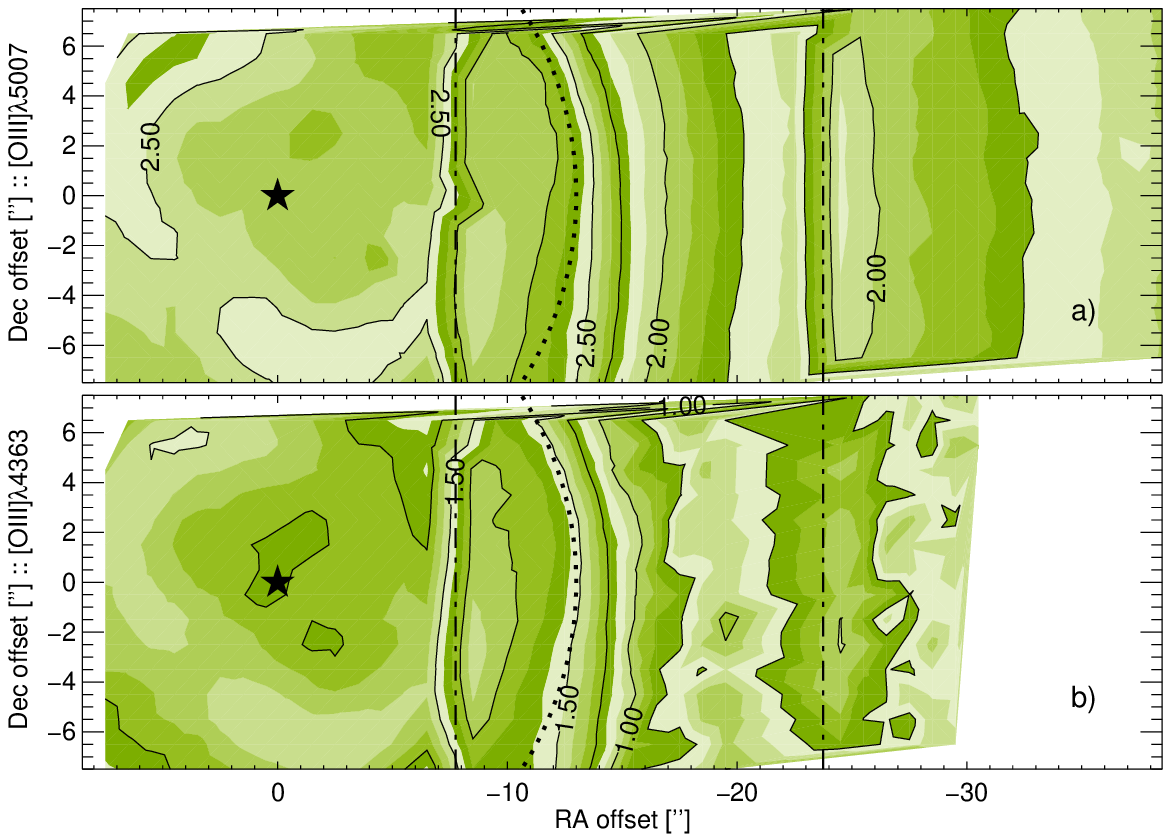}
\caption{Signal-to-noise maps of a mosaic of the three observed tiles of {\rNGCblue}, for two emission lines. In comparison to Fig.~\ref{fig:resNGC7662lmap} this figure shows the inverse relative surface brightness error for the same tiles. Values are logarithmic. From the top the two panels show: {\bf a)} [\ion{O}{iii}]$\,\lambda5007$ and {\bf b)} [\ion{O}{iii}]$\,\lambda4363$. For further details see Sect.~\ref{sec:resNGC7662sb}.}
\label{fig:resNGC7662lmape}
\end{figure*}
}

\onlfig{25}{
\begin{figure*}
  \centering
  \includegraphics{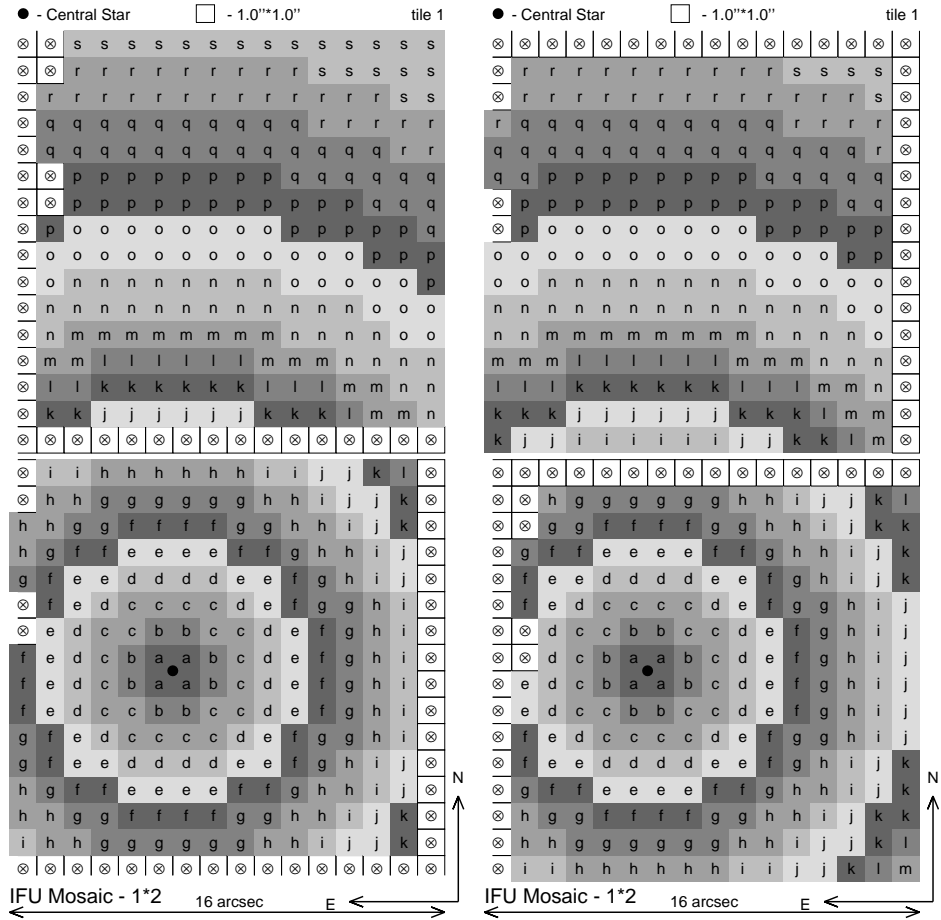}
  \caption{Binning maps used in the analysis of the observations of {\rIC}, cf.\@ Sect.~\ref{sec:resIC3568}. The map on the left hand side was used with emission lines in the blue part of the spectrum, for wavelengths shorter than $\lambda_{\text{ref}}\!=\!5\,100\,$\AA. The map on the right hand side was similarly used with emission lines of wavelengths longer than $\lambda_{\text{ref}}$. Each spaxel (square) represents an area of $1\farcs0\times1\farcs0$ on the sky. For further details of figure properties see Fig.~\ref{fig:resspxmapNGC7662}}
\label{fig:resspxmapIC3568}
\end{figure*}
}

\onlfig{26}{
\begin{figure*}
  \centering
  \includegraphics{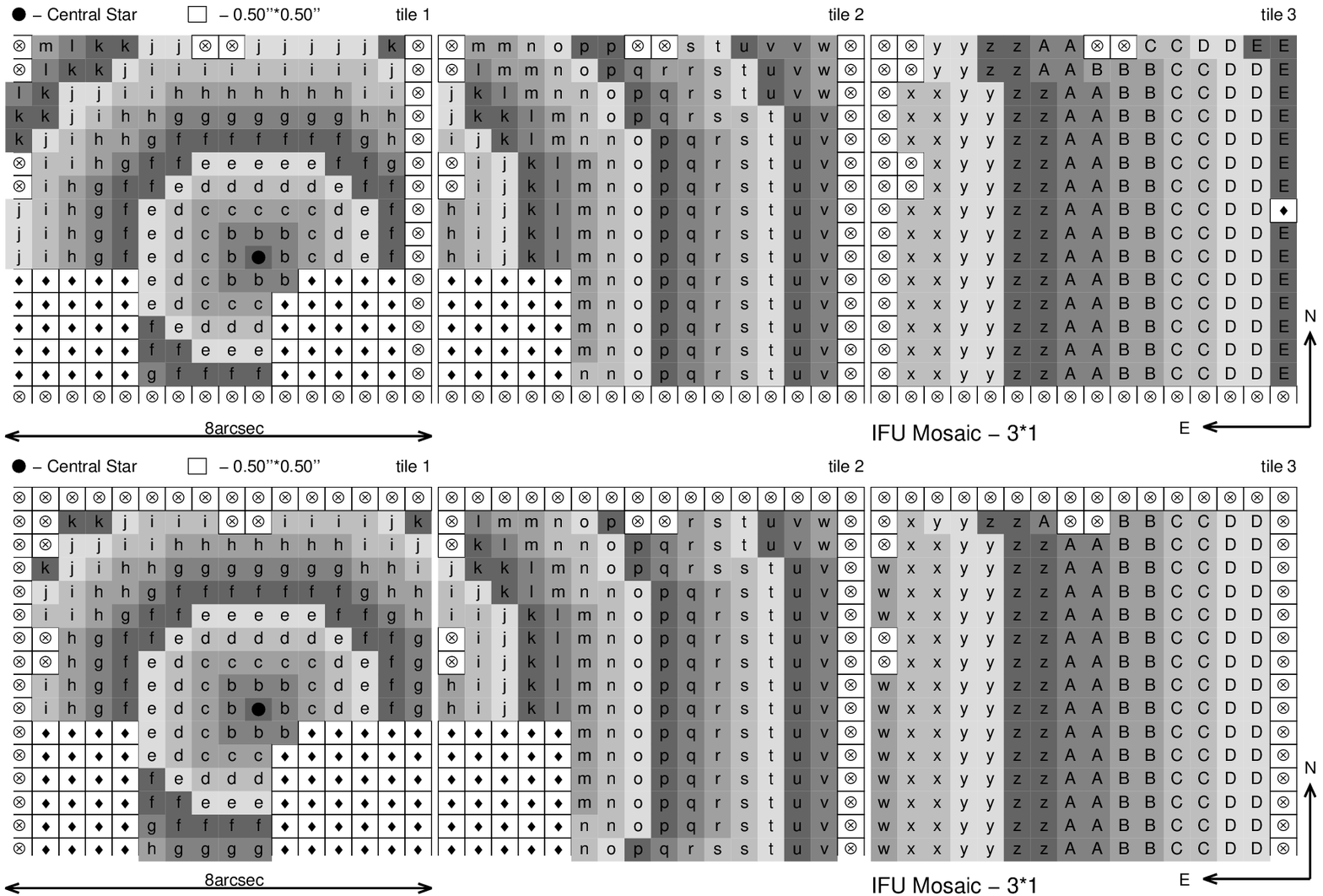}
  \caption{Binning maps used in the analysis of the observations of {\rMtt}, cf.\@ Sect.~\ref{sec:resM22}. Each spaxel (square) represents an area of $0\farcs5\times0\farcs5$ on the sky, and data is here present out to a distance of $20\arcsec$ west of the CS ($\bullet$). The map in the top panel is used with all emission lines bluewards of $\lambda_{\text{ref}}\!=\!5050\,${\AA}, and the map in the bottom panel with all emission lines redwards of this wavelength. The data in the red wavelength range was observed in a subsequent night, at a slightly different angle, which is why it requires a different pattern mask with the DAR correction. For further details of figure properties see Fig.~\ref{fig:resspxmapNGC7662}}
\label{fig:resspxmapM22}
\end{figure*}
}

\onlfig{27}{
\begin{figure*}
\centering
\includegraphics{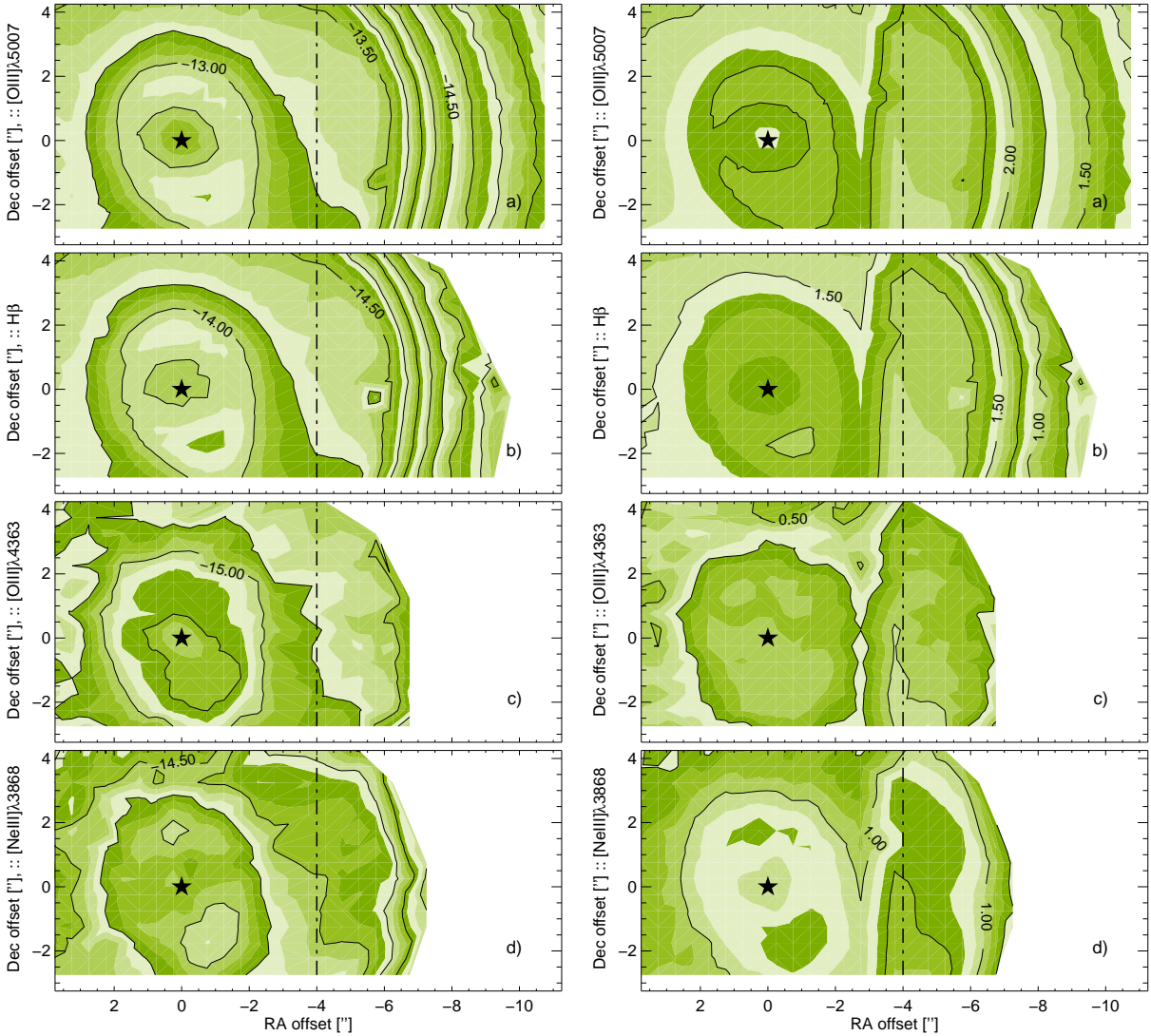}
\caption{Surface brightness maps (left) with corresponding signal-to-noise maps (right) of tiles 1 \& 2 of {\rMtt}, cf.\@ Sect.~\ref{sec:resM22sb}. From the top the panels show four emission lines of decreasing wavelength: {\bf a)} [\ion{O}{iii}]$\,\lambda5007$, {\bf b)} H$\beta$, {\bf c)} [\ion{O}{iii}]$\,\lambda4363$, and {\bf d)} [\ion{Ne}{iii}]$\,\lambda3869$. For further details of figure properties see Fig.~\ref{fig:resNGC7662lmap}. Ansae are partly visible at (RA,Dec)-offsets (--3\farcs5,--1\farcs5) and (--5,--1\farcs5). Note the low signal-to-noise of the temperature sensitive oxygen line [\ion{O}{iii}]$\,\lambda$4363 in panel {\bf c}, already at a distance of $r\simeq4\arcsec$ from the CS ($\bigstar$).}
\label{fig:resM22lmap}
\end{figure*}
}

\onlfig{28}{
\begin{figure*}
  \centering
  \includegraphics[height=18cm,angle=90]{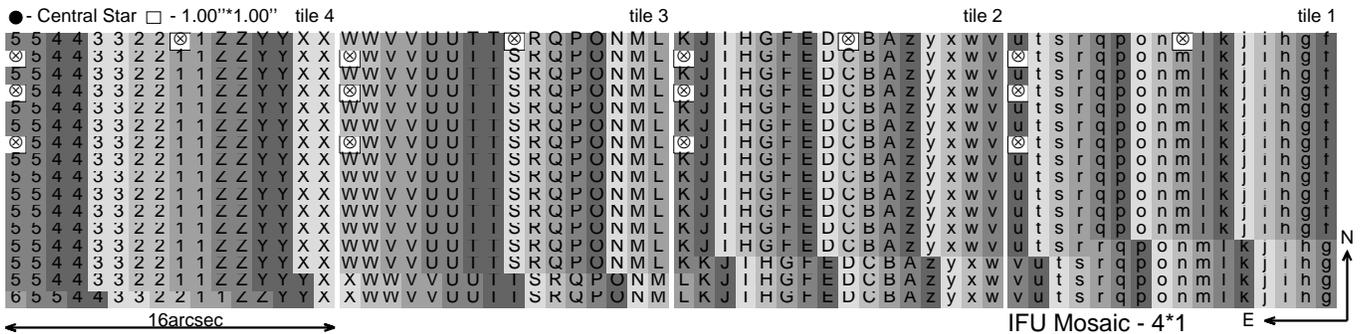}
  \caption{Binning map used in the analysis of the observations of {\rNGCowl} for the red wavelength range, cf.\@ Sect.~\ref{sec:resNGC3587}. Each spaxel (square) represents an area of $1\farcs0\!\times\!1\farcs0$ on the sky. The CS is located 66{\arcsec} to the west (right) of tile~1. For further details of figure properties see Fig.~\ref{fig:resspxmapNGC7662}}
\label{fig:resspxmapNGC3587}
\end{figure*}
}

\onlfig{29}{
\begin{figure*}
\includegraphics{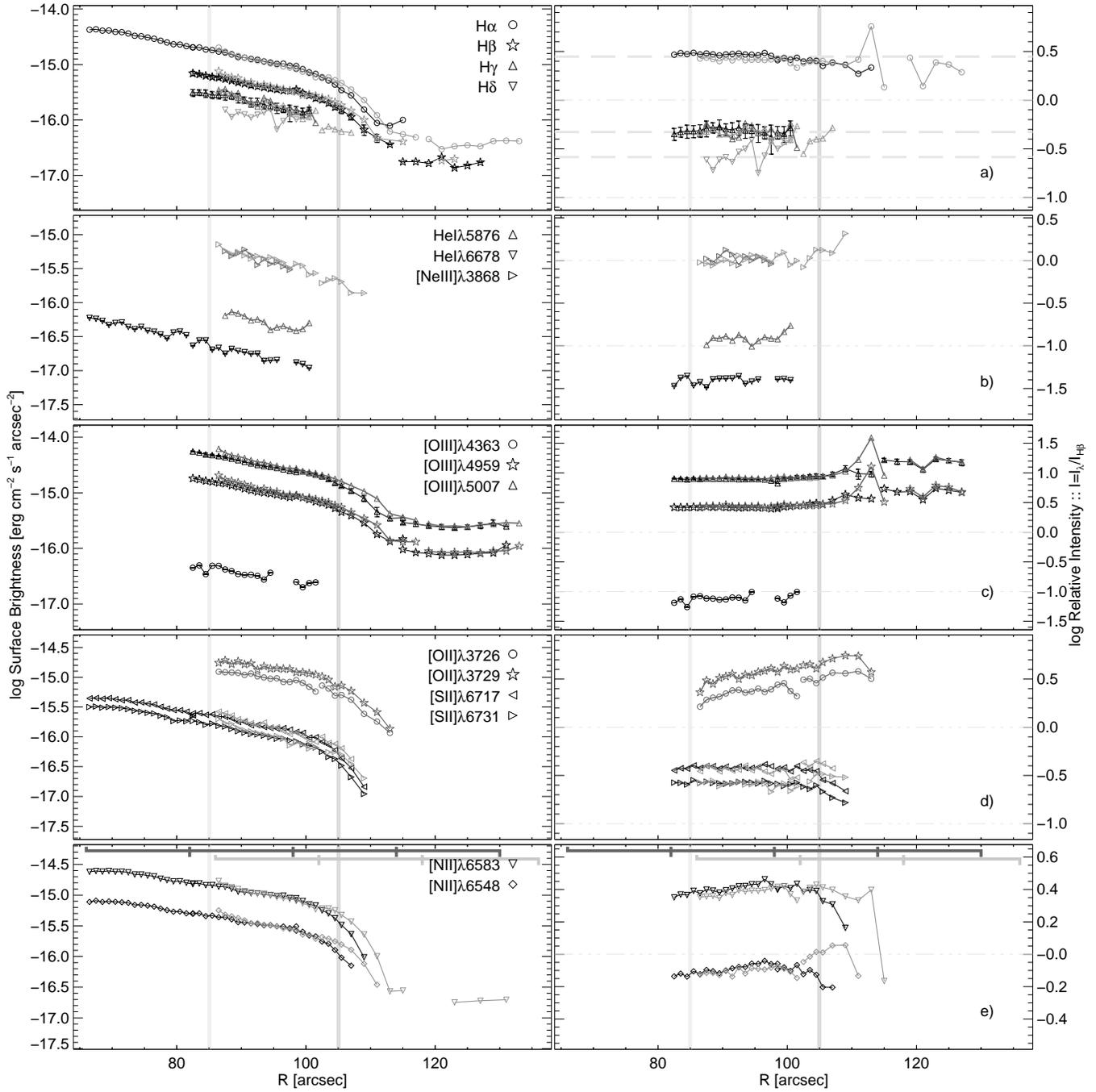}
\caption{Radial surface brightness structures of {\rNGCowl}, cf.\@ Sect.~\ref{sec:resNGC3587sb}. From the top the panels show intensities of 16 emission lines of six elements: {\bf a)} hydrogen, {\bf b)} helium and neon, {\bf c)} oxygen, {\bf d)} sulfur and oxygen, and {\bf e)} nitrogen. The locations of the rim at $r_{\text{rim}}\!=\!85\arcsec$, and the shell-halo transition at $r_{\text{shell}}\!=\!105\arcsec$, are indicated with gray vertical lines. Data of lines drawn in black were observed in April 2006, data of lines drawn in gray in February 2004. For further details of figure properties see Fig.~\ref{fig:ressbNGC7662}.}
\label{fig:ressbNGC3587}
\end{figure*}
}

\end{document}